\documentclass[prc,showpacs,nofootinbib,preprint]{revtex4-1}
\usepackage{epsfig,graphics,color}
\usepackage{subfigure}
\usepackage{bbm}
\usepackage{mathtools}
\usepackage{amssymb}
\usepackage{mathrsfs}
\usepackage{amsmath}

\newcommand{\bra}[1]{\langle #1|}
\newcommand{\ket}[1]{|#1\rangle}

\newcommand{\roundbra}[1]{(#1|}
\newcommand{\roundket}[1]{|#1)}

\def\one{{\bf 1}\,}
\def\quarknumberoperator{{\mathbbm 1}\,}

\def\tr{{\rm tr} \,}

\def\Dslash{D \hspace{-3.7mm}/}

\def\Dslash{D \hspace{-2.7mm}/ \;}

\def\vslash{v \hspace{-1.7mm}/}

\def\w2{\tilde w^2}
\def\ws2{1}

\usepackage{ulem}
\usepackage{color}

\begin{document}
\title{Combined large-$N_c$ and heavy-quark operator analysis \\for the chiral Lagrangian with charmed baryons}
\author{M.F.M. Lutz$^1$, D. Samart$^{2,4}$ and Y. Yan$^{3,4}$}
\affiliation{$^1$ GSI Helmholtzzentrum f\"ur Schwerionenforschung GmbH,\\
Planck Str. 1, 64291 Darmstadt, Germany}
\affiliation{$^2$ Rajamangala University of Technology Isan, Nakhon Ratchasima, 30000, Thailand}
\affiliation{$^3$ Suranaree University of Technology, Nakhon Ratchasima, 30000, Thailand}
\affiliation{$^4$ Thailand Center of Excellence in Physics (ThEP), Commission on Higher Education, Bangkok 10400, Thailand}
\date{\today}
\begin{abstract}
The chiral $SU(3)$ Lagrangian with charmed baryons of spin $J^P=1/2^+$ and $J^P=3/2^+$ is analyzed. We consider all
counter terms that are relevant at next-to-next-to-next-to-leading order (N$^3$LO) in a chiral extrapolation
of the charmed baryon masses. At N$^2$LO we find 16 low-energy parameters. There are 3 mass parameters for the
anti-triplet and the two sextet baryons, 6 parameters describing the meson-baryon vertices and 7 symmetry breaking
parameters. The heavy-quark spin symmetry predicts four sum rules for the meson-baryon vertices and degenerate masses
for the two baryon sextet fields. Here a large-$N_c$ operator analysis at NLO suggests the relevance of one further
spin-symmetry breaking parameter. Going from N$^2$LO to N$^3$LO adds 17 chiral symmetry breaking parameters
and 24 symmetry preserving parameters. For the leading symmetry conserving two-body counter terms
involving two baryon fields and two Goldstone boson fields we find 36 terms. While the heavy-quark spin symmetry leads to
$36-16=20$ sum rules, an expansion in $1/N_c$ at next-to-leading order (NLO) generates $36-7= 29$ parameter relations.
A combined expansion leaves 6 unknown parameters only. For the symmetry breaking counter terms we find 17 terms, for
which there are $17-9=8$ sum rules from the heavy-quark spin symmetry and $17-5=12 $ sum rules from a $1/N_c$ expansion
at NLO.
\end{abstract}

\maketitle

\section{Introduction}

The chiral $SU(3)$ Lagrangian with the charmed baryons has been used to study baryon resonances with
$J^P=\frac{1}{2}^-$ and $\frac{3}{2}^-$ quantum
numbers \cite{Lutz:2003jw,Hofmann:2005sw,Hofmann:2006qx,JimenezTejero:2009vq,Haidenbauer:2010ch,Romanets:2012ce}.
So far such studies rely on the leading order interaction characterized by the Weinberg-Tomozawa
theorem \cite{Lutz:2003jw} or they are based on phenomenological assumptions like meson-exchange
dynamics \cite{Hofmann:2005sw,Hofmann:2006qx,JimenezTejero:2009vq,Haidenbauer:2010ch,Romanets:2012ce}.
Such systems are of considerable complexity due to the possibility of charm exchange reactions like for instance
$\pi \Sigma_c \to \bar D \Sigma$. As been observed already in \cite{Hofmann:2005sw,Hofmann:2006qx} the  generated
resonance spectrum decouples approximately into two sectors. The first sector is dominated by the coupled-channel
interaction of the $D$ mesons with charm zero baryons and  the second one by the coupled-channel interaction of the
Goldstone bosons with charmed baryons. Though there are already some partial results available in the literature
it is an open challenge to derive quantitatively the spectrum of open-charm baryons in terms
of realistic interactions derived from the chiral Lagrangian.

While the leading order interaction of the Goldstone bosons with any hadron is dictated by the
chiral symmetry, this is not true for the interaction of the $D$ mesons with baryons. The $D$ mesons
interact with baryons via local counter terms that are unconstrained by chiral symmetry. Only the long-range
part of the interaction is controlled by chiral interactions, the short range part needs to be parameterized
in terms of a priori unknown contact terms. In a previous work \cite{Lutz:2011fe} such counter terms were studied
systematically in the light of the heavy-quark symmetry and large-$N_c$ sum rules. The 26 counter terms were correlated
so that only 7 unknown parameters remain. A corresponding analysis for the counter terms describing the residual
short-range interaction of the Goldstone bosons with the charmed baryons has not been performed. Such an analysis
is relevant also for an accurate flavour $SU(3)$ chiral extrapolation of the charmed ground-state baryon masses as
they are currently evaluated by various lattice QCD groups. First partial results are found in
\cite{Liu:2009jc,Bali:2012ua,Briceno:2012wt,Alexandrou:2012xk,Namekawa:2013vu,Perez-Rubio:2013oha}.

Recently it was demonstrated that a systematic and accurate flavour $SU(3)$ chiral extrapolation of the baryon octet
and decuplet states with zero charm content is feasible \cite{Semke2005,Semke2012,Semke:2012gs,Lutz:2014oxa}. Based on the chiral Lagrangian formulated
with spin 1/2 and 3/2 fields the available lattice data on the baryon masses was reproduced and accurate predictions for
the size of the low-energy parameters relevant at N$^3$LO were made. Such an analysis was made possible only by the
availability of sum rules for the low-energy parameters as they arise in the limit of a large number colors ($N_c$)
in QCD \cite{Lutz:2010se}. The latter provided a large parameter reduction that allowed fits at N$^3$LO to the lattice
data set that are significant.

The purpose of the present work is to pave the way towards a corresponding analysis of the upcoming lattice data for the
charmed baryons. In order to derive sum rules for the symmetry conserving low-energy constants we study
matrix elements of current-current correlation functions in the charmed baryon states \cite{Lutz:2010se,Lutz:2011fe}.
The technology developed in \cite{Luty1994,Dashen1994,Jenkins:1996de} will be applied.
The implications of heavy-quark symmetry on the counter terms can be worked out using a suitable multiplet
representation of the charmed baryons \cite{Yan:1992gz,Cho:1992gg,Casalbuoni:1996pg}.

The paper is organized as follows. In section II the chiral Lagrangian will be constructed.
It follows a section where the consequences of the heavy-quark spin symmetry on the low-energy constants of the
chiral Lagrangian are worked out. In section IV a corresponding large-$N_c$ operators analysis is presented.
In section V a summary of the main results is given.

\section{Chiral Lagrangian with charmed baryon fields} \label{section:chiral-lagrangian}

The construction rules for the chiral $SU(3)$ Lagrangian density are recalled.
For more technical details see for example \cite{Krause:1990xc,Yan:1992gz,Cho:1992gg,Casalbuoni:1996pg}.
The basic building blocks of the chiral Lagrangian  are
\begin{eqnarray}
&&\textcolor{black}{ U_\mu = {\textstyle{1\over 2}}\,u^\dagger \, \big(
\partial_\mu \,e^{i\,\frac{\Phi}{f}} \big)\, u^\dagger
-{\textstyle{i\over 2}}\,u^\dagger \,r_\mu\, u
+{\textstyle{i\over 2}}\,u \,l_\mu\, u^\dagger\;, \qquad
 u = e^{i\,\frac{\Phi}{2\,f}} }\,,
\nonumber\\
&&  \chi_\pm = {\textstyle{1\over 2}}\, \big(
u \,\chi_0 \,u
\pm u^\dagger \,\chi_0 \,u^\dagger
\big) \,, \qquad B_{[\bar 3]}\,, \qquad  B_{[6]}\,, \qquad B_{[6]}^\mu\,,
\label{def-fields}
\end{eqnarray}
where we include the pseudoscalar meson octet fields
$\Phi(J^P\!\!=\!0^-)$,
the baryon fields $B_{[\bar 3]}(J^P\!\!=\!{\textstyle{1\over2}}^+)$, $B_{[6]}(J^P\!\!=\!{\textstyle{1\over2}}^+)$
fields $B^\mu_{[6]}(J^P\!\!=\!{\textstyle{3\over2}}^+)$. Explicit chiral symmetry-breaking
is included in terms of scalar and pseudoscalar fields $\chi_\pm $.  They introduce the scalar and pseudo-scalar
classical source functions $s$ and $p$ with
\begin{eqnarray}
\chi_0 = 2\,B_0 \left(
\begin{array}{ccc}
m_u & 0 & 0 \\
0 &m_d & 0 \\
0 & 0& m_s \\
\end{array} \right)
 + 2\,B_0 \,( s+i\,p) \,,
\end{eqnarray}
and the quark mass matrix of QCD \cite{Gasser:1983yg,Bernard:1991zc}. The classical source functions
$r_\mu$ and $l_\mu$ in (\ref{def-fields}) are
linear combinations of the vector and axial-vector sources with $r_\mu = v_\mu+a_\mu$ and $l_\mu = v_\mu-a_\mu$.

It is convenient to decompose the fields into their isospin multiplets. The fields can be written in terms of
isospin multiplet fields
\begin{eqnarray}
&& \Phi = \tau \cdot \pi (140)
+ \alpha^\dagger \cdot  K (494) +  K^\dagger(494) \cdot \alpha
+ \eta(547)\,\lambda_8\,,
\nonumber\\
&&  \sqrt{2}\,B_{[\bar 3]}  = {\textstyle{1\over \sqrt{2}}}\,\alpha^\dagger \cdot \Xi_c(2470)
- {\textstyle{1\over \sqrt{2}}}\,\Xi_c^t(2470)\cdot \alpha
+  i\,\tau_2\,\Lambda_c(2284) \,,
\nonumber\\
&& \sqrt{2}\,B_{[6]} = {\textstyle{1\over \sqrt{2}}}\,\alpha^\dagger \cdot \Xi_c(2580)
+ {\textstyle{1\over \sqrt{2}}}\,\Xi_c^{t}(2580)\cdot \alpha
+ \Sigma_c(2453) \cdot \tau \,i\,\tau_2
\nonumber\\
&& \qquad \quad+  {\textstyle{\sqrt{2}\over 3}}\, \big(1-\sqrt{3}\,\lambda_8 \big)\,\Omega_c(2704)  \,,
\nonumber\\
&& \sqrt{2}\, B^\mu_{[6]} = {\textstyle{1\over \sqrt{2}}}\,\alpha^\dagger \cdot \Xi^{\mu}_c(2646)
+ {\textstyle{1\over \sqrt{2}}}\,\Xi_c^{t,\mu}(2646)\cdot \alpha
+ \Sigma^\mu_c(2518) \cdot \tau \,i\,\tau_2
\nonumber\\
&& \qquad \quad+  {\textstyle{\sqrt{2}\over 3}}\, \big(1-\sqrt{3}\,\lambda_8 \big)\,\Omega^{\mu}_c(2770)  \,,
\nonumber\\
&& \alpha^\dagger = {\textstyle{1\over \sqrt{2}}}\,(\lambda_4+i\,\lambda_5 ,\lambda_6+i\,\lambda_7 )\,,\qquad
\tau = (\lambda_1,\lambda_2, \lambda_3)\,,
\end{eqnarray}
where the matrices $\lambda_i$ are the standard Gell-Mann generators of the SU(3) algebra.
The numbers in the brackets recall the approximate masses of the particles in units of
MeV.

The chiral Lagrangian consists of all possible interaction
terms, formed with the fields $U^\mu,\, B_{[\bar 3]},\, B_{[6]},\, B^\mu_{[6]}  $ and $\chi_\pm$.
Derivatives of the fields must be included in compliance with the local chiral $SU(3)$ symmetry. This leads to
the notion of a covariant derivative $D_\mu$. For the flavour octet field $U_\nu$ and the flavour symmetric
sextet and flavour antisymmetric anti-triplet fields $B_{[6]}, B^\nu_{[6]}$ and $B_{[\bar 3]}$ the covariant
derivative $D_\mu$  acts as follows
\begin{eqnarray}
&&(D_\mu \, U_\nu)^{a}_{\;b} \;\,= \partial_\mu U^{a}_{\nu,b} +  \Gamma^{a}_{\mu,l}\, U^{l}_{\nu,b} -
\Gamma^{l}_{\mu,b}\, U^{a}_{\nu,l} \,, \quad
\nonumber\\
&&(D_\mu  B_{[6]})^{ab} = \partial_\mu B^{ab} +  \Gamma^{a}_{\mu,l}\, B^{lb}_{[6]} +
\Gamma^{b}_{\mu,l}\, B^{al}_{[b]} \,, \quad
\nonumber\\
&&(D_\mu  B_{[\bar 3]})^{ab} = \partial_\mu B_{[\bar 3]}^{ab} +  \Gamma^{a}_{\mu,l}\, B_{[\bar 3]}^{lb} +
\Gamma^{b}_{\mu,l}\, B_{[\bar 3]}^{al} \,,
\label{def-covariant-derivative}
\end{eqnarray}
with the chiral connection $\Gamma_\mu=-\Gamma_\mu^\dagger$ given by
\begin{eqnarray*}
&&\Gamma_\mu ={\textstyle{1\over 2}}\, u^\dagger\,
\Big[\partial_\mu -i\,r_\mu \Big] \,u
+{\textstyle{1\over 2}}\, u \,
\Big[\partial_\mu -i\,l_\mu \Big] \,u^\dagger\,.
\end{eqnarray*}

The chiral Lagrangian is a powerful tool, once it is combined with appropriate
counting rules leading to a systematic approximation strategy. We apply here
conventional power counting rules with
\begin{eqnarray}
\chi_\pm \sim Q^2 \,, \qquad U_\mu \sim Q\,, \qquad \Gamma_\mu \sim Q \,,
\end{eqnarray}
where $Q$ is a typical small scale. If a covariant derivative $D_\mu$ acts on any meson field 
it is counted as order $Q$. In contrast, owing to the large baryon masses, if a covariant derivative $D_\mu$ 
acts on a baryon mass it is counted as order $Q^0$. Nevertheless specific combinations like 
$i\,\gamma^\mu \,D_\mu - M \sim Q$ are suppressed if acting on a baryon field. Here $M$ denotes the chiral limit 
value of a baryon mass. The proper leading chiral power of a vertex involving baryon fields is 
identified upon a non-relativistic on-shell expansion, where any 3-momentum is assigned the chiral order $Q$.

In the following we construct
the leading order (LO) terms involving the three baryon fields. There are the kinetic terms
\begin{eqnarray}
&& \mathscr{L}^{(1)} \!= \tr \bar B_{[6]} \big(\gamma^\mu\, i\,D_\mu  - M^{1/2}_{[6]} \big)\,B_{[6]}
- \mathrm{tr}\,\Big(\bar{B}_{[6]}^\mu \, \big(\big[i\,\Dslash\,
-M^{3/2}_{[6]}\big]\,g_{\mu\nu} -i\,(\gamma_\mu\, D_\nu
 + \gamma_\nu \,D_\mu)
\nonumber\\
&& \qquad \;+\, \gamma_\mu\,\big[i\,\Dslash + M^{3/2}_{[6]}\big]\,\gamma_\nu \big)\,B_{[6]}^\nu\Big)
+\tr \bar B_{[\bar 3]} \big(\gamma^\mu\,i\, D_\mu  -M^{1/2}_{[\bar 3]} \big)\,B_{[\bar 3]}
\nonumber \\
&&  \qquad \;+\,F_{[66]}\,{\rm tr}\,\bar B_{[6]}\,\gamma^\mu\,\gamma_5\,i\,U_\mu\,B_{[6]}
+ F_{[\bar 3\bar 3]}\,{\rm tr}\,\bar B_{[\bar 3]}\,\gamma^\mu\,\gamma_5\,i\,U_\mu\,B_{[\bar 3]}
\nonumber\\
&&  \qquad \;+\, F_{[\bar 36]}\,{\rm tr}\,\Big( \bar B_{[6]}\,\gamma^\mu\,\gamma_5\,i\,U_\mu\,B_{[\bar 3]} + {\rm h.c.}\Big)
\nonumber\\
&&  \qquad \;+\, C_{[66]}\,{\rm tr}\,\Big( \bar B_{[6]}^\mu\,i\,U_\mu\,B^{\phantom{\mu}}_{[6]} + {\rm h.c.} \Big)
 + C_{[\bar 36]}\,{\rm tr}\,\Big( \bar B_{[6]}^\mu\,i\,U_\mu\,B^{\phantom{\mu}}_{[\bar 3]} + {\rm h.c.}\Big)
\nonumber\\
&&  \qquad \;-\, H_{[66]}\,{\rm tr}\,\bar B_{[6]}^\alpha\,g_{\alpha \beta}\,\gamma^\mu\,\gamma_5\,i\,U_\mu\,B_{[6]}^\beta \,,
\label{def-L1}
\end{eqnarray}
and 6 structures which parameterizes the three-point interactions of the Goldstone bosons with the charmed baryon
fields \cite{Yan:1992gz,Cho:1992gg}. From the kinetic terms one can read off the two-body Weinberg-Tomozawa interaction terms on which the
coupled-channel computation of \cite{Lutz:2003jw} rests. It follows upon an expansion if the kinetic terms in powers
of the Goldstone boson fields. At leading order in a chiral expansion, the bare masses
$M^{1/2}_{[6]}$, $M^{3/2}_{[6]}$ and $M^{1/2}_{[\bar 3]}$ may be identified with the flavor average of the sextet
and antitriplet baryon masses.

The main goal of this work is to derive correlations amongst the low-energy parameters introduced in the chiral Lagrangian
as they follow from a $1/N_c$ expansion. For that purpose we consider the axial-vector and scalar currents,
\begin{eqnarray}
&& A_\mu^{(a)}(x) = \bar \Psi (x)\,\gamma_\mu \,\gamma_5\,\frac{\lambda_a}{2}\,\Psi(x) \,, \qquad \qquad
S^{(a)}(x) = \bar \Psi (x)\,\frac{\lambda_a}{2}\,\Psi(x) \,,
\label{def-amu}
\end{eqnarray}
in baryon matrix elements, where we recall their definitions in terms of the
Heisenberg quark-field operators $\Psi(x)$. With $\lambda_a$ we
denote the Gell-Mann matrices supplemented with a singlet matrix $\lambda_0 = \sqrt{2/3}\,\one $.
Such matrix elements can be analyzed systematically in the $1/N_c$
expansion \cite{Dashen1994,Jenkins:1996de}. Given the chiral Lagrangian, it is well defined how to derive the
contribution of a given term to such matrix elements. The classical matrices of source functions, $a_\mu(x)$ and $s(x)$,
enter the chiral Lagrangian via the building block
\begin{eqnarray}
U_\mu =  \frac{i}{2\,f}\,\partial_\mu \,\Phi - i\,a_\mu + \cdots \,, \qquad
\chi_+ = 2\,B_0\, s + \cdots \,,
\label{result:Umu}
\end{eqnarray}
where for notational simplicity in the following we put $B_0 = 1/2$.

For our matching purposes it suffices to take matrix elements in the flavour $SU(3)$ limit where the physical baryon states
\begin{eqnarray}
\ket{p,\, ij_\pm ,\,S,\,\chi }\,,
\label{def-states}
\end{eqnarray}
are specified by the momentum $p$ and the flavor indices $i,j,k=1,2,3$. The spin $S$ and the spin-polarization
are $\chi = 1,2$ for the spin one-half ($S=1/2$) and $\chi =1,\cdots ,4$ for the spin three-half states $(S=3/2)$. The
flavour sextet and the anti-triplet are discriminated by their symmetric (index $+$) and anti-symmetric (index $-$)
behaviour under the exchange of $i \leftrightarrow j$.

From the leading order chiral Lagrangian (\ref{def-L1})
the matrix elements of the axial-vector current are readily computed at tree-level
\begin{eqnarray}
&& \bra{\bar p\,,\,mn_+,\, {\textstyle{1\over 2}},\,\bar\chi\,}\,A_i^{(a)}(0)\,
\ket{\,p\,,kl_+, \, {\textstyle{1\over 2}},\,\chi }
= \frac12\,\sigma_{\bar\chi\chi}^{(i)}\,F_{[66]}\,\Lambda_{(kl)_+}^{(a)\,,\,(mn)_+} +\cdots \,,
\nonumber\\
&& \bra{\bar p\,,\,mn_-,\, {\textstyle{1\over 2}},\,\bar\chi\,}\,A_i^{(a)}(0)\,
\ket{\,p\,,kl_-, \, {\textstyle{1\over 2}},\,\chi }
= \frac12\,\sigma_{\bar\chi\chi}^{(i)}\,F_{[\bar 3\bar 3]}\,\Lambda_{(kl)_-}^{(a)\,,\,(mn)_-}+\cdots \,,
\nonumber\\
&& \bra{\bar p\,,\,mn_+,\, {\textstyle{1\over 2}},\,\bar\chi\,}\,A_i^{(a)}(0)\,
\ket{\,p\,,kl_-, \, {\textstyle{1\over 2}},\,\chi }
=  \frac12\,\sigma_{\bar\chi\chi}^{(i)}\,F_{[\bar 36]}\,\Lambda_{(kl)_-}^{(a)\,,\,(mn)_+}+\cdots \,,
\nonumber\\
&& \bra{\bar p\,,\,mn_+,\, {\textstyle{3\over 2}},\,\bar\chi\,}\,A_i^{(a)}(0)\,
\ket{\,p\,,kl_+, \, {\textstyle{1\over 2}},\,\chi }
= \textcolor{black}{+}\,\frac12\,S_{\bar\chi\chi}^{(i)}\,C_{[66]}\,\Lambda_{(kl)_+}^{(a)\,,\,(mn)_+}+\cdots \,,
\nonumber\\
&& \bra{\bar p\,,\,mn_+,\, {\textstyle{3\over 2}},\,\bar\chi\,}\,A_i^{(a)}(0)\,
\ket{\,p\,,kl_-, \, {\textstyle{1\over 2}},\,\chi }
= \textcolor{black}{+}\,\frac12 \,S_{\bar\chi\chi}^{(i)}\,C_{[\bar 36]}\,\Lambda_{(kl)_-}^{(a)\,,\,(mn)_+}+\cdots \,,
\nonumber\\
&& \bra{\bar p\,,\,mn_+,\, {\textstyle{3\over 2}},\,\bar\chi\,}\,A_i^{(a)}(0)\,
\ket{\,p\,,kl_+, \, {\textstyle{3\over 2}},\,\chi }
= \frac12 \,\big(\vec S\,\sigma^{(i)}\,\vec S^\dagger\big)_{\bar\chi\chi}\,H_{[66]}\,\Lambda_{(kl)_+}^{(a)\,,\,(mn)_+}+\cdots \,,
\label{axial-current element non-relativistic}
\end{eqnarray}
where we introduced convenient flavour and spin structures
\begin{eqnarray}
&&\Lambda_{(ij)_\pm}^{(a),\,(mn)_\pm}
=  \frac 14\,\Big( \lambda_{mi}^{(a)}\,\delta_{nj} \pm \lambda_{ni}^{(a)}\,\delta_{mj}
\pm \lambda_{mj}^{(a)}\,\delta_{ni} + \lambda_{nj}^{(a)}\,\delta_{mi} \,\Big)\,,
\nonumber\\
&& \Lambda_{(ij)_\pm}^{(a),\,(mn)_\mp}
=  \frac 14\,\Big( \lambda_{mi}^{(a)}\,\delta_{nj} \mp \lambda_{ni}^{(a)}\,\delta_{mj}
\pm \lambda_{mj}^{(a)}\,\delta_{ni} - \lambda_{nj}^{(a)}\,\delta_{mi} \,\Big)\,,
\nonumber\\ \nonumber\\
&& S^\dagger_i\, S_j= \delta_{ij} - \frac{1}{3}\sigma_i \sigma_j \,, \qquad
S_i\,\sigma_j - S_j\,\sigma_i = -i\,\varepsilon_{ijk} \,S_k\,,
\qquad \vec S\cdot   \vec S^\dagger= \one_{(4\times 4)}\,,
\nonumber\\
&& \vec S^\dagger \cdot  \vec S =2\, \one_{(2\times 2)}\,, \qquad \vec S \cdot \vec \sigma = 0 \,,\qquad
\epsilon_{ijk}\,S_i\,S^\dagger_j = i\,\vec S \,\sigma_k\,\vec S^\dagger\,.
\label{def:spin-transition-matrices}
\end{eqnarray}
The dots in (\ref{scalar-current element non-relativistic}) represent additional terms that are suppressed
as the three momenta  $\bar p$ and $p$ approach zero. Such correction terms are not relevant for our
matching purposes. Here we also assumed degenerate baryon masses for the baryon states
as they arise in the large-$N_c$ limit. The well known result (\ref{def:spin-transition-matrices}) settles
our conventions and notations and is also conveniently matched to the results of a large-$N_c$ operator analysis.
For the latter its suffices to study the spatial components of the axial-vector current only.

At next-to-leading order (NLO) there are symmetry conserving and symmetry breaking terms.  We identify 7
symmetry breaking counter terms
\allowdisplaybreaks[1]
\begin{eqnarray}
&&\mathscr{L}^{(2)}_\chi =  b_{1,[\bar 3\bar 3]}\,{\rm tr}\,\big( \bar B_{[\bar 3]}\,B_{[\bar 3]}\big)\,{\rm tr}\,\big(\chi_+\big)
+b_{2,[\bar 3\bar 3]}\,{\rm tr}\,\big( \bar B_{[\bar 3]}\,\chi_+\,B_{[\bar 3]}\big)
+b_{1,[\bar 3 6]}\,{\rm tr}\,\big( \bar B_{[6]}\,\chi_+\,B_{[\bar 3]} + {\rm h.c.}\big)
\nonumber\\
&& \qquad \;\,+\,
b_{1,[66]}\,{\rm tr}\,\big( \bar B_{[6]}\,B_{[6]}\big)\,{\rm tr}\,\big(\chi_+\big)
+ b_{2,[66]}\,{\rm tr}\,\big( \bar B_{[6]}\,\chi_+\,B_{[6]}\,\big)
\nonumber\\
&& \qquad \;\,-\, d_{1,[66]}\,{\rm tr}\,\big(  g_{\mu \nu}\,\bar B_{[6]}^\mu\,B_{[6]}^\nu\big)\,{\rm tr}\,\big(\chi_+\big)
- d_{2,[66]}\,{\rm tr}\,\big( g_{\mu \nu}\,\bar B_{[6]}^\mu\,\chi_+\,B_{[6]}^\nu\,\big)\,.
\label{def-L2-chi}
\end{eqnarray}
From the leading order chiral Lagrangian (\ref{def-L2-chi})
the matrix elements of the scalar current are readily computed at tree-level. Here we consider the
singlet $(a=0)$ and octet $(a=1, \cdots , 8)$ components with
\begin{eqnarray}
&& \bra{\bar p\,,\,mn_+,\, {\textstyle{1\over 2}},\,\bar\chi\,}\,S^{(a)}(0)\,
\ket{\,p\,,kl_+, \, {\textstyle{1\over 2}},\,\chi }
= -\,\Big(  \sqrt{\textstyle{ 3 \over  2}}\,b_{1,[6 6]}\,\delta_{a0}\,\delta_{(kl)_+}^{(mn)_+}
+ { \textstyle {1 \over 2}}\,b_{2,[6 6]}\,\Lambda_{(kl)_+}^{(a)\,,\,(mn)_+}\Big) \,\delta_{\bar\chi\chi} \,,
\nonumber\\
&& \bra{\bar p\,,\,mn_-,\, {\textstyle{1\over 2}},\,\bar\chi\,}\,S^{(a)}(0)\,
\ket{\,p\,,kl_-, \, {\textstyle{1\over 2}},\,\chi }
=  -\,\Big( \delta_{a0}\, \sqrt{\textstyle{ 3 \over  2}}\,b_{1,[\bar 3 \bar 3]}\,\delta_{(kl)_-}^{(mn)_-}
+ { \textstyle {1 \over 2}}\,b_{2,[\bar 3\bar 3]}\,\Lambda_{(kl)_-}^{(a)\,,\,(mn)_-}\Big) \,\delta_{\bar\chi\chi}\,,
\nonumber\\
&& \bra{\bar p\,,\,mn_-,\, {\textstyle{1\over 2}},\,\bar\chi\,}\,S^{(a)}(0)\,
\ket{\,p\,,kl_+, \, {\textstyle{1\over 2}},\,\chi }
= -{ \textstyle {1 \over 2}}\,\,b_{1,[\bar 3\bar 6]}\,\Lambda_{(kl)_+}^{(a)\,,\,(mn)_-}\,\delta_{\bar\chi\chi} \,,
\nonumber\\
&& \bra{\bar p\,,\,mn_+,\, {\textstyle{3\over 2}},\,\bar\chi\,}\,S^{(a)}(0)\,
\ket{\,p\,,kl_+, \, {\textstyle{3\over 2}},\,\chi }
=-\,\Big( \sqrt{\textstyle{ 3 \over  2}}\, d_{1,[6 6]}\,\delta_{a0}\,\delta_{(kl)_+}^{(mn)_+}
+{ \textstyle {1 \over 2}}\,d_{2,[6 6]}\, \Lambda_{(kl)_-}^{(a)\,,\,(mn)_+}\Big)\,\delta_{\bar\chi\chi}  \,
\nonumber\\
&& \bra{\bar p\,,\,mn_+,\, {\textstyle{3\over 2}},\,\bar\chi\,}\,S^{(a)}(0)\,
\ket{\,p\,,kl_+, \, {\textstyle{1\over 2}},\,\chi }
= 0 \,,
\nonumber\\
&& \bra{\bar p\,,\,mn_+,\, {\textstyle{3\over 2}},\,\bar\chi\,}\,S^{(a)}(0)\,
\ket{\,p\,,kl_-, \, {\textstyle{1\over 2}},\,\chi }
= 0\,,
\label{scalar-current element non-relativistic}
\end{eqnarray}
where we use the notations introduced in (\ref{def:spin-transition-matrices}) with
\begin{eqnarray}
&& \lambda_0 = \sqrt{\frac 2 3}\,\one \,, \qquad \qquad \delta_{(ij)_\pm}^{(mn)_\pm}
=  \frac 12\,\Big( \delta_{mi}\,\delta_{nj} \pm \delta_{ni}\,\delta_{mj} \,\Big)\,.
\label{def-one-pm}
\end{eqnarray}
Like in  (\ref{axial-current element non-relativistic}) additional terms that are not relevant in this work are not shown in (\ref{scalar-current element non-relativistic}). Such terms vanish in the limit $p, \bar p \to \to 0$.

There are 36 symmetry preserving terms counter terms in $\mathscr{L}^{(2)}$.
Following \cite{Lutz2002a,Lutz:2010se}
the symmetry conserving counter term are classified according to their Dirac structure.
\allowdisplaybreaks[1]
\begin{eqnarray}
&&\mathscr{L}^{(2)} = \mathscr{L}^{(2)}_\chi +\mathscr{L}^{(S)} + \mathscr{L}^{(V)} + \mathscr{L}^{(A)} + \mathscr{L}^{(T)}\,.
\label{def-L2}
\end{eqnarray}
A complete list relevant at second order is
\begin{eqnarray}
&& \mathscr{L}^{(S)} \!= g_{0,[\bar 3\bar 3]}^{(S)}\,{\rm tr}\,\big( \bar B_{[\bar 3]}\,B_{[\bar 3]}\big)\,{\rm tr}\,\big( U_\mu\,U^\mu\big)
+ g_{D,[\bar 3\bar 3]}^{(S)}\,{\rm tr}\,\big( \bar B_{[\bar 3]}\,\big\{ U_\mu,\,U^\mu\big\}\, B_{[\bar 3]}\big)
\nonumber\\
&& \qquad \;+\,
g_{0,[66]}^{(S)}\,{\rm tr}\,\big( \bar B_{[6]}\,B_{[6]}\big)\,{\rm tr}\,\big( U_\mu\,U^\mu\big)
+ g_{1,[66]}^{(S)}\,{\rm tr}\,\big( \bar B_{[6]}\, U^\mu\,B_{[6]}\, U^T_\mu\big)
\nonumber\\
&& \qquad \;+\, g_{D,[66]}^{(S)}\,{\rm tr}\,\big( \bar B_{[6]}\,\big\{ U_\mu,\,U^\mu\big\}\, B_{[6]}\big)
+ g_{1,[\bar 3 6]}^{(S)}\,{\rm tr}\,\big( \bar B_{[6]}\, U^\mu\,B_{[\bar 3]}\, U^T_\mu + {\rm h.c.}\big)
\nonumber\\
&& \qquad \;+\,g_{D,[\bar 36]}^{(S)}\,{\rm tr}\,\big( \bar B_{[6]}\,\big\{ U_\mu,\,U^\mu\big\}\, B_{[\bar 3]} + {\rm h.c.}\big)
\nonumber\\
&& \qquad \;+\, h_{0,[66]}^{(S)}\,{\rm tr}\,\big( \bar B_{[6]}^\mu\,g_{\mu \nu}\,B_{[6]}^\nu\big)\,{\rm tr}\,\big( U_\alpha\,U^\alpha\big)
+ h_{1,[66]}^{(S)}\,{\rm tr}\,\big( \bar B_{[6]}^\mu\,B_{[6]}^\nu\big)\,{\rm tr}\,\big( U_\mu\,U_\nu\big)
\nonumber\\
&& \qquad \;+\, h_{2,[66]}^{(S)}\,{\rm tr}\,\big( \bar B_{[6]}^\mu\, g_{\mu \nu}\,\big\{ U^\alpha,\,U_\alpha\big\}\, B_{[6]}^\nu\big)
+ h_{3,[66]}^{(S)}\,{\rm tr}\,\big( \bar B_{[6]}^\mu\,\big\{ U_\mu,\,U_\nu\big\}\, B_{[6]}^\nu\big)
\nonumber\\
&& \qquad \;+\, h_{4,[66]}^{(S)}\,{\rm tr}\,\big( \bar B_{[6]}^\mu\,g_{\mu \nu}\,U^\alpha\, B_{[6]}^\nu\,U^T_\alpha \big)
+ h_{5,[66]}^{(S)}\,{\rm tr}\,\big( \bar B_{[6]}^\mu\,U_\nu\, B_{[6]}^\nu\,U^T_\mu + \bar B_{[6]}^\mu\,U_\mu\, B_{[6]}^\nu\,U^T_\nu\big)\,,
\nonumber\\
&& \mathscr{L}^{(V)} \!= g_{0,[\bar 3\bar 3]}^{(V)}\,{\rm tr}\,\big( \bar B_{[\bar 3]}\,i\,\gamma^\alpha\,(D^\beta B_{[\bar 3]})\,{\rm tr}\,\big( U_\beta\,U_\alpha\big) + {\rm h.c.} \big)
\nonumber\\
&& \qquad \;+\, g_{1,[\bar 3\bar 3]}^{(V)}\,{\rm tr}\,\big( \bar B_{[\bar 3]}\,i\,\gamma^\alpha\,U_\beta\,(D^\beta B_{[\bar 3]})\, U^T_\alpha + \bar B_{[\bar 3]}\,i\,\gamma^\alpha\,U_\alpha\,(D^\beta B_{[\bar 3]})\, U^T_\beta + {\rm h.c.}\big)
\nonumber\\
&& \qquad \;+\,
 g_{D,[\bar 3\bar 3]}^{(V)}\,{\rm tr}\,\big( \bar B_{[\bar 3]}\,i\,\gamma^\alpha\,\big\{ U_\alpha,\,U_\beta\big\}\,(D^\beta  B_{[\bar 3]}) + {\rm h.c.} \big)
\nonumber\\
&& \qquad \;+\,  g_{1,[\bar 36]}^{(V)}\,{\rm tr}\,\big( \bar B_{[6]}\,i\,\gamma^\alpha\, U_\alpha\,(D^\beta  B_{[\bar 3]})\,U^T_\beta
+ \bar B_{[6]}\,i\,\gamma^\alpha\, U_\beta\,(D^\beta  B_{[\bar 3]})\,U^T_\alpha
\nonumber\\
&& \qquad \qquad \quad \;\;\, \,-\,
\textcolor{black}{
(D^\beta\bar B_{[6]})\,i\,\gamma^\alpha\, U_\alpha\,  B_{[\bar 3]}\,U^T_\beta
- (D^\beta \bar B_{[6]})\,i\,\gamma^\alpha\, U_\beta\,B_{[\bar 3]}\,U^T_\alpha
}
 + {\rm h.c.} \big)
\nonumber\\
&& \qquad \;+\,  g_{D,[\bar 36]}^{(V)}\,{\rm tr}\,\big( \bar B_{[6]}\,i\,\gamma^\alpha\,\big\{ U_\alpha,\,U_\beta\big\}\,(D^\beta  B_{[\bar 3]})
- (D^\beta \bar B_{[6]})\,i\,\gamma^\alpha\,\big\{ U_\alpha,\,U_\beta\big\}\, B_{[\bar 3]}
+ {\rm h.c.} \big)
\nonumber\\
&& \qquad \;+\,  g_{0,[66]}^{(V)}\left(\,{\rm tr}\,\big( \bar B_{[6]}\,i\,\gamma^\alpha\,(D^\beta B_{[6]})\big)\,{\rm tr}\,\big( U_\beta\,U_\alpha\big) + {\rm h.c.} \right)
\nonumber\\
&& \qquad \;+\, g_{1,[66]}^{(V)}\,{\rm tr}\,\big( \bar B_{[6]}\,i\,\gamma^\alpha\,U_\beta\,(D^\beta B_{[6]})\, U^T_\alpha + \bar B_{[6]}\,i\,\gamma^\alpha\,U_\alpha\,(D^\beta B_{[6]})\, U^T_\beta + {\rm h.c.}\big)
\nonumber\\
&& \qquad \;+\, g_{D,[66]}^{(V)}\,{\rm tr}\,\big( \bar B_{[6]}\,i\,\gamma^\alpha\,\big\{ U_\alpha,\,U_\beta\big\}\,(D^\beta  B_{[6]}) + {\rm h.c.}\big)
\nonumber\\
&& \qquad \;+\, h_{0,[66]}^{(V)}\,{\rm tr}\,\big( \bar B_{[6]}^\mu\,g_{\mu \nu}\,i\,\gamma^\alpha\,(D^\beta B_{[6]}^\nu)\,{\rm tr}\,\big( U_\alpha\,U_\beta\big) + {\rm h.c.} \big)
\nonumber\\
&& \qquad \;+\,  h_{1,[66]}^{(V)}\,{\rm tr}\,\big( \bar B_{[6]}^\mu\,g_{\mu \nu}\, i\,\gamma^\alpha\,U_\beta\,(D^\beta B_{[6]}^\nu)\, U^T_\alpha + \bar B_{[6]}^\mu\,g_{\mu \nu}\,i\,\gamma^\alpha\,U_\alpha\,(D^\beta B_{[6]}^\nu)\, U^T_\beta + {\rm h.c.}\big)
\nonumber\\
&& \qquad \;+\, h_{2,[66]}^{(V)}\,{\rm tr}\,\big( \bar B^\mu_{[6]}\,g_{\mu \nu}\,i\,\gamma^\alpha\,\big\{ U_\alpha,\,U_\beta\big\}\,(D^\beta B_{[6]}^\nu ) + {\rm h.c.}\big)  \,,
\nonumber\\
&&\mathscr{L}^{(A)} \!= f_{0,[66]}^{(A)}\,{\rm tr}\,\big( \bar B_{[6]}^\mu\,\gamma^\nu\,\gamma_5\,B_{[6]}\,{\rm tr}\,\big( U_\nu\,U_\mu\big) + {\rm h.c.} \big)
\nonumber\\
&& \qquad \;+\, f_{1,[66]}^{(A)}\,{\rm tr}\,\big( \bar B_{[6]}^\mu\,\gamma^\nu\,\gamma_5\,U_\nu\,B_{[6]}\,U^T_\mu
+ \bar B_{[6]}^\mu\,\gamma^\nu\,\gamma_5\,U_\mu\,B_{[6]}\,U^T_\nu + {\rm h.c.} \big)
\nonumber\\
&& \qquad \;+\,f_{D,[66]}^{(A)}\,{\rm tr}\,\big( \bar B_{[6]}^\mu\,\gamma^\nu\,\gamma_5\,\big\{ U_\mu,\,U_\nu\big\}\, B_{[6]} + {\rm h.c.}\big)
+ f_{F,[66]}^{(A)}\,{\rm tr}\,\big( \bar B_{[6]}^\mu\,\gamma^\nu\,\gamma_5\,\big[ U_\mu,\,U_\nu\big]\, B_{[6]} + {\rm h.c.}\big)
\nonumber\\
&& \qquad \;+\, f_{1,[\bar 36]}^{(A)}\,{\rm tr}\,\big( \bar B_{[6]}^\mu\,\gamma^\nu\,\gamma_5\,U_\nu\,B_{[\bar 3]}\,U^T_\mu
\textcolor{black}{-} \bar B_{[6]}^\mu\,\gamma^\nu\,\gamma_5\,U_\mu\,B_{[\bar 3]}\,U^T_\nu + {\rm h.c.} \big)
\nonumber\\
&& \qquad \;+\, f_{D,[\bar 36]}^{(A)}\,{\rm tr}\,\big( \bar B_{[6]}^\mu\,\gamma^\nu\,\gamma_5\,\big\{ U_\mu,\,U_\nu\big\}\, B_{[\bar 3]} + {\rm h.c.}\big)
+ f_{F,[\bar 36]}^{(A)}\,{\rm tr}\,\big( \bar B_{[6]}^\mu\,\gamma^\nu\,\gamma_5\,\big[ U_\mu,\,U_\nu\big]\, B_{[\bar 3]} + {\rm h.c.}\big) \,,
\nonumber\\
&&\mathscr{L}^{(T)} \!= g_{F,[\bar 3\bar 3]}^{(T)}\,{\rm tr}\,\big( \bar B_{[\bar 3]}\,i\,\sigma^{\alpha \beta}\,\big[ U_\alpha,\,U_\beta\big]\, B_{[\bar 3]}\big)
+ g_{1,[\bar 36]}^{(T)}\,{\rm tr}\,\big( \bar B_{[6]}\,i\,\sigma^{\alpha \beta}\,U_\alpha\, B_{[\bar 3]}\,U^T_\beta + {\rm h.c.}\big)
\nonumber\\
&& \qquad \;+\, g_{F,[\bar 36]}^{(T)}\,{\rm tr}\,\big( \bar B_{[6]}\,i\,\sigma^{\alpha \beta}\,\big[ U_\alpha,\,U_\beta\big]\, B_{[\bar 3]} + {\rm h.c.}\big)
+ g_{F,[66]}^{(T)}\,{\rm tr}\,\big( \bar B_{[6]}\,i\,\sigma^{\alpha \beta}\,\big[ U_\alpha,\,U_\beta\big]\, B_{[6]}\big)
\nonumber\\
&& \qquad \;+\, h_{F,[66]}^{(T)}\,{\rm tr}\,\big( \bar B^\mu_{[6]}\,g_{\mu \nu}\,i\,\sigma^{\alpha \beta }\,\big[ U_\alpha,\,U_\beta \big]\, B^\nu_{[6]}\big)\,,
\nonumber
\end{eqnarray}
where further possible terms that are redundant owing to the  on-shell conditions
of spin-$\frac32$ fields with $\gamma_\mu\,B_{[6]}^\mu = 0$ and $\partial_\mu\,B_{[6]}^\mu = 0$ are
eliminated systematically \footnote{Note that the term $g_{1,[\bar 3 \bar 3]}^{(V)}$ is redundant at leading order in a 
non-relativistic expansion. Its leading effect is recovered by the particular choice 
$g_{D,[\bar 3 \bar 3]}^{(V)}= - 2\,g_{0,[\bar 3 \bar 3]}^{(V)} = 2\,g_{1,[\bar 3 \bar 3]}^{(V)}$. }.

The symmetry conserving parameters contribute to the current-current correlation
function of two time-ordered axial-vector currents
\begin{eqnarray}
&& A^{\,ab}_{\mu \nu} (q) = i\,\int d^4 x \,e^{-i\,q\cdot x}  \,{\mathcal T}\,A^{(a)}_\mu (x)\,A^{(b)}_\nu(0) \,,
\label{def:axial-axial}
\end{eqnarray}
in the baryon states. The latter can be analyzed systematically in the $1/N_c$ expansion \cite{Lutz:2010se,Lutz:2011fe}.
In order to prepare for a matching we derive the specific form of such contributions
\begin{eqnarray}
&& \bra{\bar p,\,mn_+,\, {\textstyle{1\over 2}},\,\bar\chi\,}\,A_{ij}^{ab}(q)\,
\ket{p,\,kl_+,\, {\textstyle{1\over 2}},\,\chi\,} =
\epsilon_{ijk}\,\sigma_{\bar\chi\chi}^{(k)}\,g_{F,[66]}^{(T)}\,f_{abc}\,\Lambda_{(kl)_+}^{(c),\,(mn)_+}
\nonumber\\
&& \qquad +\,\delta_{ij}\,\delta_{\bar\chi\,\chi}\,\frac 12\,
\Big\{ \Big(g_{0,[66]}^{(S)} \textcolor{black}{+ \frac 2 3 \,g_{D,[66]}^{(S)}}- \frac 13\,g_{1,[66]}^{(S)}\Big)\,2\,\delta_{ab}\,\delta_{(kl)_+}^{(mn)_+}
\nonumber\\
&& \qquad \qquad \quad +\, \Big(g_{D,[66]}^{(S)} -  \frac 12\,g_{1,[66]}^{(S)}\Big)\,2\,d_{abc}\,\Lambda_{(kl)_+}^{(c),\,(mn)_+}
\nonumber\\
&& \qquad \qquad \quad +\, g_{1,[66]}^{(S)}\,\Big(\Lambda_{(rs)_+}^{(a),\,(mn)_+}\,\Lambda_{(kl)_+}^{(b),\,(rs)_+} + \Lambda_{(rs)_+}^{(b),\,(mn)_+}\,\Lambda_{(kl)_+}^{(a),\,(rs)_+} \Big) \Big\}
\nonumber\\
&&\qquad -\,\frac{(\bar p + p)_i\,(\bar p + p)_j }{M}\,\delta_{\bar\chi\,\chi}\,
\Big\{ \Big(g_{0,[66]}^{(V)} \textcolor{black}{+  \frac 2 3\,g_{D,[66]}^{(V)}}- \frac 23\,g_{1,[66]}^{(V)}\Big)\,\delta_{ab}\,\delta_{(kl)_+}^{(mn)_+}
\nonumber\\
&& \qquad \qquad \quad +\, \Big(g_{D,[66]}^{(V)} - g_{1,[66]}^{(V)}\Big)\,d_{abc}\,\Lambda_{(kl)_+}^{(c),\,(mn)_+}
\nonumber\\
&& \qquad \qquad \quad +\, g_{1,[66]}^{(V)}\,\Big(\Lambda_{(rs)_+}^{(a),\,(mn)_+}\,\Lambda_{(kl)_+}^{(b),\,(rs)_+} + \Lambda_{(rs)_+}^{(b),\,(mn)_+}\,\Lambda_{(kl)_+}^{(a),\,(rs)_+} \Big) \Big\}
 + \cdots\,,
\nonumber\\
\nonumber\\
&&\bra{\bar p,\,mn_-,\, {\textstyle{1\over 2}},\,\bar\chi\,}\,A_{ij}^{ab}(q)\,
\ket{p,\,kl_-,\, {\textstyle{1\over 2}},\,\chi\,} = \epsilon_{ijk}\,\sigma_{\bar\chi\chi}^{(k)}\,g_{F,[\bar3\bar3]}^{(T)}
\,f_{abc}\,\Lambda_{(kl)_-}^{(c),\,(mn)_-}
\nonumber\\
&&\qquad -\,\frac{(\bar p + p)_i\,(\bar p + p)_j}{M}\,\delta_{\bar\chi\,\chi}\,
\Big\{ \Big(g_{0,[\bar3\bar3]}^{(V)} \textcolor{black}{+\frac 23\,g_{D,[\bar3\bar3]}^{(V)}}\,\textcolor{black}{+} \frac 23\,g_{1,[\bar3\bar3]}^{(V)}\Big)\,\delta_{ab}\,\delta_{(kl)_-}^{(mn)_-}
\nonumber\\
&& \qquad \qquad \quad +\, \Big(g_{D,[\bar3\bar3]}^{(V)} \textcolor{black}{+} g_{1,[\bar3\bar3]}^{(V)}\Big)\,d_{abc}\,\Lambda_{(kl)_-}^{(c),\,(mn)_-}
\nonumber\\
&& \qquad \qquad \quad \textcolor{black}{-}\, g_{1,[\bar3\bar3]}^{(V)}\,\Big(\Lambda_{(rs)_-}^{(a),\,(mn)_-}\,\Lambda_{(kl)_-}^{(b),\,(rs)_-} + \Lambda_{(rs)_-}^{(b),\,(mn)_-}\,\Lambda_{(kl)_-}^{(a),\,(rs)_-} \Big) \Big\}
\nonumber\\
&&\qquad +\,\delta_{ij}\,\delta_{\bar\chi\,\chi}\,\frac 12\,
\Big\{ 2\,\Big( g_{0,[\bar3\bar3]}^{(S)} \textcolor{black}{+ \frac 2 3\,g_{D,[\bar3\bar3]}^{(S)}}\Big) \,\delta_{ab}\,\delta_{(kl)_-}^{(mn)_-}
 + 2\,g_{D,[\bar3\bar3]}^{(S)}\,d_{abc}\,\Lambda_{(kl)_-}^{(c),\,(mn)_-} \Big\} + \cdots\,,
\nonumber\\
\nonumber\\
&&\bra{\bar p,\,mn_-,\, {\textstyle{1\over 2}},\,\bar\chi\,}\,A_{ij}^{ab}(q)\,
\ket{p,\,kl_+,\, {\textstyle{1\over 2}},\,\chi\,} =
\epsilon_{ijk}\,\sigma_{\bar\chi\chi}^{(k)}\,\Big\{
\Big( g_{F,[\bar36]}^{(T)} - \frac12\,g_{1,[\bar36]}^{(T)}\Big)\,f_{abc}\,\Lambda_{(kl)_+}^{(c),\,(mn)_-}
\nonumber\\
&& \qquad \qquad \quad \textcolor{black}{-}\,\frac i2\,g_{1,[\bar36]}^{(T)}\,\Big(\Lambda_{(rs)_+}^{(a),\,(mn)_-}\,\Lambda_{(kl)_+}^{(b),\,(rs)_+} - \Lambda_{(rs)_+}^{(b),\,(mn)_-}\,\Lambda_{(kl)_+}^{(a),\,(rs)_+} \Big)\Big\}
\nonumber\\
&&\qquad -\,\frac{(\bar p + p)_i\,(\bar p + p)_j}{M}\,\delta_{\bar\chi\,\chi}\,
\Big\{\Big(\textcolor{black}{2}\,g_{D,[\bar36]}^{(V)} \textcolor{black}{+} g_{1,[\bar36]}^{(V)}\Big)\,d_{abc}\,\Lambda_{(kl)_+}^{(c),\,(mn)_-}
\nonumber\\
&& \qquad \qquad \quad \textcolor{black}{-}\, g_{1,[\bar36]}^{(V)}\,\Big(\Lambda_{(rs)_+}^{(a),\,(mn)_-}\,\Lambda_{(kl)_+}^{(b),\,(rs)_+} + \Lambda_{(rs)_+}^{(b),\,(mn)_-}\,\Lambda_{(kl)_+}^{(a),\,(rs)_+} \Big) \Big\}
\nonumber\\
&&\qquad +\,\delta_{ij}\,\delta_{\bar\chi\,\chi}\,\Big\{\Big(g_{D,[\bar36]}^{(S)} \textcolor{black}{+} \frac 12\,g_{1,[\bar36]}^{(S)}\Big)\,d_{abc}\,\Lambda_{(kl)_+}^{(c),\,(mn)_-}
\nonumber\\
&& \qquad \qquad \quad \textcolor{black}{-}\, \frac12\,g_{1,[\bar36]}^{(S)}\,\Big(\Lambda_{(rs)_+}^{(a),\,(mn)_-}\,\Lambda_{(kl)_+}^{(b),\,(rs)_+} + \Lambda_{(rs)_+}^{(b),\,(mn)_-}\,\Lambda_{(kl)_+}^{(a),\,(rs)_+} \Big) \Big\} + \cdots\,,
\nonumber\\
\nonumber\\
&& \bra{\bar p,\,mn_+,\, {\textstyle{3\over 2}},\,\bar\chi\,}\,A_{ij}^{ab}(q)\,
\ket{p,\,kl_+,\, {\textstyle{1\over 2}},\,\chi\,} =
i\,\epsilon_{ijk}\,S_{\bar\chi\chi}^{(k)} \,\frac 12\,f_{F,[66]}^{(A)}\,i\,f_{abc}\,\Lambda_{(kl)_+}^{(c),\,(mn)_+}
\nonumber\\
&&\qquad  -\,\Big( S_i\,\sigma_j + S_j\,\sigma_i \Big)_{\bar\chi\chi}
\,\frac 12\,\Big\{ \Big(f_{0,[66]}^{(A)}\textcolor{black}{+ \frac 2 3 \,f_{D,[66]}^{(A)}} - \frac 23\,f_{1,[66]}^{(A)}\Big)\,\delta_{ab}\,\delta_{(kl)_+}^{(mn)_+}
\nonumber\\
&&\qquad \qquad \quad +\, \Big(f_{D,[66]}^{(A)} - f_{1,[66]}^{(A)}\Big)\,d_{abc}\,\Lambda_{(kl)_+}^{(c),\,(mn)_+}
\nonumber\\
&&\qquad \qquad \quad +\, f_{1,[66]}^{(A)}\,\Big(\Lambda_{(rs)_+}^{(a),\,(mn)_+}\,\Lambda_{(kl)_+}^{(b),\,(rs)_+} + \Lambda_{(rs)_+}^{(b),\,(mn)_+}\,\Lambda_{(kl)_+}^{(a),\,(rs)_+} \Big) \Big\} + \cdots\,,
\nonumber\\
\nonumber\\
&& \bra{\bar p,\,mn_+,\, {\textstyle{3\over 2}},\,\bar\chi\,}\,A_{ij}^{ab}(q)\,
\ket{p,\,kl_-,\, {\textstyle{1\over 2}},\,\chi\,} =
\frac{i}{2}\,\epsilon_{ijk}\,S_{\bar\chi\chi}^{(k)} \,\Big\{
 \Big( f_{F,[\bar36]}^{(A)} + f_{1,[\bar36]}^{(A)} \Big) \,i\,f_{abc}\,\Lambda_{(kl)_-}^{(c),\,(mn)_+}
\nonumber\\
&&\qquad  -\, f_{1,[\bar36]}^{(A)}\,\Big(\Lambda_{(rs)_+}^{(a),\,(mn)_+}\,\Lambda_{(kl)_-}^{(b),\,(rs)_+} 
- \Lambda_{(rs)_+}^{(b),\,(mn)_+}\,\Lambda_{(kl)_-}^{(a),\,(rs)_+} \Big)\Big\} + \cdots\,,
\nonumber\\ \nonumber\\
&& \bra{\bar p,\,mn_+,\, {\textstyle{3\over 2}},\,\bar\chi\,}\,A_{ij}^{ab}(q)\,
\ket{p,\,kl_+,\, {\textstyle{3\over 2}},\,\chi\,} =
\textcolor{black}{-}\,\epsilon_{ijk}\,\big(\vec S\,\sigma^{(k)}\,\vec S^\dagger\big)_{\bar\chi\chi}\,h_{F,[66]}^{(T)}\,f_{abc}\,\Lambda_{(kl)_+}^{(c),\,(mn)_+}
\nonumber\\
&&\qquad -\, \delta_{ij}\,\delta_{\bar\chi\,\chi}\,\frac 12\,
\Big\{ \Big(h_{0,[66]}^{(S)}\textcolor{black}{+ \frac 23\,h_{2,[66]}^{(S)}} - \frac 13\,h_{4,[66]}^{(S)}\Big)\,2\,\delta_{ab}\,\delta_{(kl)_+}^{(mn)_+}
\nonumber\\
&&\qquad \qquad \quad  +\, \Big(h_{2,[66]}^{(S)} - \frac 12\,h_{4,[66]}^{(S)}\Big)\,2\,d_{abc}\,\Lambda_{(kl)_+}^{(c),\,(mn)_+}
\nonumber\\
&&\qquad \qquad \quad  +\, h_{4,[66]}^{(S)}\,\Big(\Lambda_{(rs)_+}^{(a),\,(mn)_+}\,\Lambda_{(kl)_+}^{(b),\,(rs)_+} + \Lambda_{(rs)_+}^{(b),\,(mn)_+}\,\Lambda_{(kl)_+}^{(a),\,(rs)_+} \Big) \Big\}
\nonumber\\
&&\qquad -\,\Big( S_i\,S_j^\dagger + S_j\,S_i^\dagger\Big)_{\bar\chi\,\chi}\,
\frac 12\,\Big\{ \Big(h_{1,[66]}^{(S)} \textcolor{black}{+ \frac 2 3\,h_{3,[66]}^{(S)}} - \frac 23\,h_{5,[66]}^{(S)}\Big)\,\delta_{ab}\,\delta_{(kl)_+}^{(mn)_+}
\nonumber\\
&&\qquad \qquad \quad  +\, \Big(h_{3,[66]}^{(S)} - h_{5,[66]}^{(S)}\Big)\,d_{abc}\,\Lambda_{(kl)_+}^{(c),\,(mn)_+}
\nonumber\\
&&\qquad \qquad \quad +\, h_{5,[66]}^{(S)}\,\Big(\Lambda_{(rs)_+}^{(a),\,(mn)_+}\,\Lambda_{(kl)_+}^{(b),\,(rs)_+} + \Lambda_{(rs)_+}^{(b),\,(mn)_+}\,\Lambda_{(kl)_+}^{(a),\,(rs)_+} \Big) \Big\}
\nonumber\\
&& \qquad \textcolor{black}{+}\,\frac{(\bar p + p)_i\,(\bar p + p)_j}{M}\,\delta_{\bar\chi\,\chi}
\Big\{ \Big(h_{0,[66]}^{(V)} \textcolor{black}{+ \frac 23\,h_{2,[66]}^{(V)}} - \frac 23\,h_{1,[66]}^{(V)}\Big)\,\delta_{ab}\,\delta_{(kl)_+}^{(mn)_+}
\nonumber\\
&&\qquad \qquad \quad  +\, \Big(h_{2,[66]}^{(V)} - h_{1,[66]}^{(V)}\Big)\,d_{abc}\,\Lambda_{(kl)_+}^{(c),\,(mn)_+}
\nonumber\\
&& \qquad \qquad \quad +\, h_{1,[66]}^{(V)}\,\Big(\Lambda_{(rs)_+}^{(a),\,(mn)_+}\,\Lambda_{(kl)_+}^{(b),\,(rs)_+} + \Lambda_{(rs)_+}^{(b),\,(mn)_+}\,\Lambda_{(kl)_+}^{(a),\,(rs)_+} \Big) \Big\}
+ \cdots \,,
\label{non-relativistic-correlator}
\end{eqnarray}
where $q= \bar p-p$ and $a,b =1, \cdots 8$. For the spin and flavour structures we apply our notations
(\ref{def:spin-transition-matrices}, \ref{def-one-pm}). Like in (\ref{axial-current element non-relativistic}) the dots in (\ref{non-relativistic-correlator})
represent additional terms that are further suppressed for small 3-momenta $p$ and $\bar p$.
Here we also assumed a degenerate baryon mass $M$ for the baryon states as they arise in the large-$N_c$ limit.
Like in (\ref{axial-current element non-relativistic}) we focus on the spatial components of the axial-vector
currents as they suffice to establish the desired correlations of the counter terms that arise in the
large-$N_c$ limit.

We close this section with a collection of the terms in 
$\mathscr{L}^{(4)}$ that are relevant in a chiral extrapolation of the baryon masses at N$^3$LO. 
The symmetry conserving counter terms in $\mathscr{L}^{(4)}$ do not contribute to the
two point function at tree-level. Therefore they contribute to the baryon self energies
at N$^5$LO. There are 17 symmetry breaking counter terms
\begin{eqnarray}
&& \mathscr{L}_{\chi}^{(4)} \!= c_{1,[\bar 3\bar 3]}\,{\rm tr}\,\big( \bar B_{[\bar 3]}\,B_{[\bar 3]}\big)\,{\rm tr}\,\big(\chi_+^2\big)
+ c_{2,[\bar 3\bar 3]}\,{\rm tr}\,\big( \bar B_{[\bar 3]}\,B_{[\bar 3]}\big)\,\big({\rm tr}\,\chi_+\big)^2
\nonumber\\
&& \qquad \;+\, c_{3,[\bar 3\bar 3]}\,{\rm tr}\,\big( \bar B_{[\bar 3]}\,\chi_+\,B_{[\bar 3]}\,\big)\,{\rm tr}\,\big(\chi_+\big)
+  c_{4,[\bar3\bar3]}\,{\rm tr}\,\big( \bar B_{[\bar3]}\,\chi_+^2\,B_{[\bar3]}\big)
\nonumber\\
&& \qquad \; + \,c_{1,[66]}\,{\rm tr}\,\big( \bar B_{[6]}\,B_{[6]}\big)\,{\rm tr}\,\big(\chi_+^2\big)
+ c_{2,[66]}\,{\rm tr}\,\big( \bar B_{[6]}\,B_{[6]}\big)\,\big({\rm tr}\,\chi_+\big)^2
\nonumber\\
&& \qquad \;+\, c_{3,[66]}\,{\rm tr}\,\big( \bar B_{[6]}\,\chi_+\,B_{[6]}\,\big)\,{\rm tr}\,\big(\chi_+\big)
+ c_{4,[66]}\,{\rm tr}\,\big( \bar B_{[6]}\,\chi_+^2\, B_{[6]}\big)
+ c_{5,[66]}\,{\rm tr}\,\big( \bar B_{[6]}\,\chi_+\, B_{[6]}\,\chi^T_+\big)
\nonumber\\
&& \qquad \;+\, c_{1,[\bar 3 6]}\,{\rm tr}\,\big( \bar B_{[6]}\,\chi_+\,B_{[\bar 3]}+ {\rm h.c.}\,\big)\,{\rm tr}\,\big(\chi_+\big)
+  c_{2,[\bar3 6]}\,{\rm tr}\,\big( \bar B_{[6]}\,\chi_+^2\,B_{[\bar3]}+ {\rm h.c.}\big)
\nonumber\\
&& \qquad \;+\, c_{3,[\bar3 6]}\,{\rm tr}\,\big( \bar B_{[6]}\,\chi_+\, B_{[\bar3]}\,\chi^T_+ + {\rm h.c.}\big)
\nonumber\\
&& \qquad \; -\, e_{1,[66]}\,{\rm tr}\,\big( \bar B_{[6]\mu}\,B_{[6]}^\mu\big)\,{\rm tr}\,\big(\chi_+^2\big)
- e_{2,[66]}\,{\rm tr}\,\big( \bar B_{[6]\mu}\,B_{[6]}^\mu\big)\,\big({\rm tr}\,\chi_+\big)^2
\nonumber\\
&& \qquad \;-\, e_{3,[66]}\,{\rm tr}\,\big( \bar B_{[6]\mu}\,\chi_+\,B_{[6]}^\mu\,\big)\,{\rm tr}\,\big(\chi_+\big)
- e_{4,[66]}\,{\rm tr}\,\big( \bar B_{[6]\mu}\,\chi_+^2\, B_{[6]}^\mu\big)
\nonumber\\
&& \qquad \;-\, e_{5,[66]}\,{\rm tr}\,\big( \bar B_{[6]\mu}\,\chi_+\, B_{[6]}^\mu\,\chi^{\textcolor{black}{T}}_+\big)\,.
\label{def-L4}
\end{eqnarray}
The symmetry breaking counter terms contribute to the current-current correlation
function of two time-ordered scalar currents
\begin{eqnarray}
&& S^{ab} (q) = i\,\int d^4 x \,e^{-i\,q\cdot x}  \,{\mathcal T}\,S^{(a)} (x)\,S^{(b)}(0) \,,
\label{def:scalar-scalar}
\end{eqnarray}
in the baryon states. Like in (\ref{scalar-current element non-relativistic}) we consider singlet and octet
components with $a,b = 0, \cdots 8$.
The latter can be analyzed systematically in the $1/N_c$ expansion \cite{Lutz:2010se,Lutz:2011fe}.
In order to prepare for a matching we derive the specific form of such contributions
\begin{eqnarray}
&& \bra{\bar p,\,mn_+,\, {\textstyle{1\over 2}},\,\bar\chi\,}\,S^{ab}(q)\,
\ket{p,\,kl_+,\, {\textstyle{1\over 2}},\,\chi\,} =
\textcolor{black}{+}\,\delta_{\bar\chi\,\chi}\,\frac 12\,
\Big\{ \Big(c_{1,[66]} \textcolor{black}{ +\frac 13\,c_{4,[66]}}- \frac 13\,c_{5,[66]}\Big)\,2\,\delta_{ab}\,\delta_{(kl)_+}^{(mn)_+}
\nonumber\\
&& \qquad \quad +\,\textcolor{black}{ \Big(c_{4,[66]} - c_{5,[66]}\Big)\,}d_{abc}\,\Lambda_{(kl)_+}^{(c),\,(mn)_+}
 +c_{5,[66]}\,\Big(\Lambda_{(rs)_+}^{(a),\,(mn)_+}\,\Lambda_{(kl)_+}^{(b),\,(rs)_+} + \Lambda_{(rs)_+}^{(b),\,(mn)_+}\,\Lambda_{(kl)_+}^{(a),\,(rs)_+} \Big)
\nonumber\\
&& \qquad \quad +\, c_{2,[66]}\,6\,\delta_{a0}\,\delta_{b0}\,\delta_{(kl)_+}^{(mn)_+} + c_{3,[66]}\,\sqrt{\frac32}\,\Big(\delta_{a0}\,\Lambda_{(kl)_+}^{(b),\,(mn)_+} + \delta_{b0}\,\Lambda_{(kl)_+}^{(a),\,(mn)_+} \Big) \Big\} + \cdots\,,
\nonumber\\
\nonumber\\
&&\bra{\bar p,\,mn_-,\, {\textstyle{1\over 2}},\,\bar\chi\,}\,S^{ab}(q)\,
\ket{p,\,kl_-,\, {\textstyle{1\over 2}},\,\chi\,} =
\textcolor{black}{+}\,\delta_{\bar\chi\,\chi}\,\frac 12\,
\Big\{ \Big( c_{1,[\bar3\bar3]} + \textcolor{black}{ \frac 1 3 c_{4,[\bar3\bar3]}}\Big)\,2\,\delta_{ab}\,\delta_{(kl)_-}^{(mn)_-}
\nonumber\\
&& \qquad \quad  +\, \textcolor{black}{c_{4,[\bar3\bar3]} }\,d_{abc}\,\Lambda_{(kl)_-}^{(c),\,(mn)_-}
+ c_{2,[\bar3\bar3]}\,6\,\delta_{a0}\,\delta_{b0}\,\delta_{(kl)_-}^{(mn)_-}
\nonumber\\
&& \qquad \quad  +\,c_{3,[\bar3\bar3]}\,\sqrt{\frac32}\,\Big(\delta_{a0}\,\Lambda_{(kl)_-}^{(b),\,(mn)_-} + \delta_{b0}\,\Lambda_{(kl)_-}^{(a),\,(mn)_-} \Big) \Big\} + \cdots\,,
\nonumber\\
\nonumber\\
&&\bra{\bar p,\,mn_-,\, {\textstyle{1\over 2}},\,\bar\chi\,}\,S^{ab}(q)\,
\ket{p,\,kl_+,\, {\textstyle{1\over 2}},\,\chi\,} =
\textcolor{black}{+}\,\delta_{\bar\chi\,\chi}\,\frac 12\,
\Big\{ \textcolor{black}{\Big(c_{2,[\bar36]} - c_{3,[\bar36]}\Big)}\,d_{abc}\,\Lambda_{(kl)_+}^{(c),\,(mn)_-}
\nonumber\\
&& \qquad \quad +\, c_{3,[\bar36]}\,\Big(\Lambda_{(rs)_-}^{(a),\,(mn)_-}\,\Lambda_{(kl)_+}^{(b),\,(rs)_-} + \Lambda_{(rs)_-}^{(b),\,(mn)_-}\,\Lambda_{(kl)_+}^{(a),\,(rs)_-} \Big)
\nonumber\\
&& \qquad \quad +\, c_{1,[\bar36]}\,\sqrt{\frac32}\,\Big(\delta_{a0}\,\Lambda_{(kl)_+}^{(b),\,(mn)_-} + \delta_{b0}\,\Lambda_{(kl)_+}^{(a),\,(mn)_-} \Big) \Big\} + \cdots\,,
\nonumber\\
\nonumber\\
&& \bra{\bar p,\,mn_+,\, {\textstyle{3\over 2}},\,\bar\chi\,}\,S^{ab}(q)\,
\ket{p,\,kl_\pm,\, {\textstyle{1\over 2}},\,\chi\,} = 0 + \cdots\,,
\nonumber\\
\nonumber\\
&& \bra{\bar p,\,mn_+,\, {\textstyle{3\over 2}},\,\bar\chi\,}\,S^{ab}(q)\,
\ket{p,\,kl_+,\, {\textstyle{3\over 2}},\,\chi\,} =
\textcolor{black}{+}\,\delta_{\bar\chi\,\chi}\,\frac 12\,
\Big\{ \Big(e_{1,[66]} \textcolor{black}{+ \frac 13\,e_{4,[66]}}- \frac 13\,e_{5,[66]}\Big)\,2\,\delta_{ab}\,\delta_{(kl)_+}^{(mn)_+}
\nonumber\\
&& \qquad \quad +\,  \textcolor{black}{\Big(e_{4,[66]} - e_{5,[66]}\Big)}\,d_{abc}\,\Lambda_{(kl)_+}^{(c),\,(mn)_+}
 + e_{5,[66]}\,\Big(\Lambda_{(rs)_+}^{(a),\,(mn)_+}\,\Lambda_{(kl)_+}^{(b),\,(rs)_+} + \Lambda_{(rs)_+}^{(b),\,(mn)_+}\,\Lambda_{(kl)_+}^{(a),\,(rs)_+} \Big)
\nonumber\\
&& \qquad \quad +\, e_{2,[66]}\,6\,\delta_{a0}\,\delta_{b0}\,\delta_{(kl)_+}^{(mn)_+} + e_{3,[66]}\,\sqrt{\frac32}\,\Big(\delta_{a0}\,\Lambda_{(kl)_+}^{(b),\,(mn)_+} + \delta_{b0}\,\Lambda_{(kl)_+}^{(a),\,(mn)_+} \Big) \Big\} + \cdots \,,
\label{non-relativistic-correlator-SS}
\end{eqnarray}
where the intermediate flavour index $c= 1, \cdots , 8$ is summed over the octet components only. We use the convention
\begin{eqnarray}
d_{0ab}=d_{a0b}=d_{ab0}=\sqrt{\frac 2 3}\,\delta_{ab}\,.
\label{def-d0ab}
\end{eqnarray}

Altogether we introduced $9+7+36+17= 69$ distinct low-energy constants, which have to be determined.
Out of the 69 terms there are 56 distinct terms relevant in a chiral expansion of the baryon masses at N$^3$LO.
This may appear a hopeless situation. However, this is not the case due to the heavy-quark spin symmetry and sum rules that
arise in the large-$N_c$ limit of QCD.

\newpage

\section{Heavy quark mass expansion}\label{section:HQS}

In the limit of an infinite charm quark mass the two sextet fields can be combined into a super
multiplet field \cite{Georgi:1990cx,Yan:1992gz,Cho:1992gg,Casalbuoni:1996pg}. This reflects the fact that in this
limit the $\frac{1}{2}^+$ and $\frac{3}{2}^+$ baryons are related by a spin flip of the charm quark, which
does not cost any energy. Therefore, the properties of such states are closely related.
In order to work out the implications of the heavy-quark symmetry of QCD it is useful to introduce auxiliary and
slowly varying fields, $B_\pm(x)$ and $B^\mu_\pm(x)$. We decompose the baryon sextet fields into such components
\begin{eqnarray}
&& B_{[6]}(x) \;\;\,\,\!= e^{-i\,(v\cdot x) \,M^{1/2}_{[6]}}\,B_{+}(x) +e^{+i\,(v\cdot x) \,M^{1/2}_{[6]}}\,B_{-}(x)\,,
\nonumber\\
&& B^\mu_{[6]}(x) \;\;\,\,\!= e^{-i\,(v\cdot x) \,M^{3/2}_{[6]}}\,B^\mu_{+}(x) +e^{+i\,(v\cdot x) \,M^{3/2}_{[6]}}\,B^\mu_{-}(x)\,,
\label{non-relativistic-expansion}
\end{eqnarray}
with a 4-velocity $v$ normalized by $v^2=1$. The mass parameters $M^{1/2}_{[6]}$ and  $M^{3/2}_{[6]}$ are the chiral limit
masses of the two sextet baryons as introduced in (\ref{def-L1}). A corresponding decomposition for the anti-triplet field
$B_{[\bar 3]}(x)$ is assumed. As a consequence of Eq.~(\ref{non-relativistic-expansion}), time and
spatial derivatives of the fields $B_\pm$ and $B^\mu_\pm$ are small compared to $M_{[6]}\,v_\alpha\,B_\pm(x)$. In the limit
$M_{[6]} \to \infty$ the former terms can be neglected. Note that the fields $B_\pm(x)$ and $B_\pm^\mu(x)$
annihilate quanta with charm content $\pm 1$.

The mass parameters $M^{1/2}_{[\bar 3]}$, $M^{1/2}_{[6]}$ and $M^{3/2}_{[6]}$ may be expanded in inverse powers of the
charm quark mass $M_c$. A matching with QCD's properties \cite{Georgi:1990cx,Yan:1992gz,Cho:1992gg,Jenkins:1996de}
leads to the scaling properties
\begin{eqnarray}
M^{3/2}_{[6]}- M^{1/2}_{[6]} \sim \frac{1}{M_c} \,, \qquad \qquad M^{1/2}_{[6]} - M^{1/2}_{[\bar 3]} \sim 1\,,
\label{larg-Mc-scalaing}
\end{eqnarray}
which implies that the two sextet masses are degenerate in the heavy-quark mass limit. We apply here the formalism developed
in \cite{Georgi:1990cx,Yan:1992gz,Cho:1992gg,Casalbuoni:1996pg} and
introduce the multiplet field $H^\mu_{[6]}$, connected to the fields $B_+(x)$ and $B_+^\mu(x)$ in the heavy-quark mass limit
as follows\footnote{Note that
${\tr } \gamma_5\,\gamma_\mu \,\gamma_\nu \,\gamma_\alpha \,\gamma_\beta =
-4\,i\,\epsilon_{\mu \nu \alpha \beta }
$
in the convention used in this work.}
\begin{eqnarray}
&& H^\mu_{[6]} = \frac{1}{\sqrt 3}\,(\gamma^\mu + v^\mu)\,\gamma_5\,\frac{1+ \vslash}{2}\,B_{+} + \frac{1+ \vslash}{2}\,B_{+}^\mu\,,\qquad
 \bar H^\mu_{[6]} =  \big(H^\mu_{[6]}\big)^\dagger\,\gamma_0 \,,
\label{baryon-supermultiplet}
\end{eqnarray}
According to \cite{Georgi:1990cx,Cho:1992gg}, the field $H^\mu_{[6]}$ transforms
under the heavy-quark spin symmetry group $SU_v(2)$, the elements of which being
characterized by the 4-vector $\theta^\alpha$ with $\theta \cdot v=0$, as follows:
\begin{eqnarray}
&&H^\mu_{[6]} \to  e^{-i\,J_\alpha\,\theta^\alpha}\,H^\mu_{[6]}
\,, \qquad \quad
\bar H^\mu_{[6]} \to  \big(e^{-i\,J_\alpha\,\theta^\alpha}\,H^\mu_{[6]}\big)^\dagger\,\gamma_0 =
\bar H^\mu_{[6]}\,e^{+i\,J_\alpha\,\theta^\alpha}\,,
\nonumber\\
&& J_\alpha =\frac{1}{2}\,\gamma_5\, [\vslash, \gamma_\alpha] \,,
\qquad \qquad \,J^\dagger_\alpha \, \gamma_0=\gamma_0\,J_\alpha \,, \qquad \;\;
[\vslash, J_\alpha ]_- = 0 \,.
\label{spin-rotation}
\end{eqnarray}
In (\ref{spin-rotation}) we denote the heavy-quark spin-operator with $J_\alpha$ as to avoid
any notational confusion with the spin transition matrices $S_i$ introduced in (\ref{def:spin-transition-matrices}).
Under a Lorentz transformation $\Lambda_{\mu \nu}$, characterized by the antisymmetric tensor $\omega_{\mu \nu}$,
the spinor part of the field transforms as
\begin{eqnarray}
&& H^\mu_{[6]} \to e^{+i\,J_{\alpha \beta}\,\omega^{\alpha \beta}}\,\Lambda^{\mu}_{\; \nu}\,H^\nu_{[6]}\,,
\qquad \qquad  J_{\alpha \beta} = \frac{i}{4}\,[\gamma_\alpha, \gamma_\beta]\,,
\nonumber\\
&&\bar H^\mu_{[6]} \to \Lambda^{\mu}_{\; \nu}\,\bar H^\nu_{[6]}\,e^{-i\,J_{\alpha \beta}\,\omega^{\alpha \beta}}\,,
\end{eqnarray}
An analogous construction is applied for the flavour anti-triplet field
\begin{eqnarray}
&& B_{[\bar 3]}(x)= e^{-i\,(v\cdot x) \,M_c}\,B^{(+)}_{[\bar 3]}(x) +e^{+i\,(v\cdot x) \,M_c}\,B^{(-)}_{[\bar 3]}(x)\,,
\nonumber\\
&& H_{[\bar 3]} = \frac{1+ \vslash}{2}\,B^{(+)}_{[\bar 3]}\,, \qquad
\bar H_{[\bar 3]} = \left( H_{[\bar 3]} \right)^\dagger\,\gamma_0 \,.
\end{eqnarray}
It follows that only field combinations of the form, where there is a nontrivial Dirac matrix  neither left of the
fields $H^\mu_{[6]}$ and $H^{\phantom{\mu}}_{\bar 3}$ nor right to the fields $\bar H^\mu_{[6]}$ and  $\bar H^{\phantom{\mu}}_{[\bar 3]}$,
are invariant under the spin group $SU_v(2)$. With this rule it is straightforward to construct $SU_v(2)$-invariant
interaction terms. There is one exception to that rule. The $\gamma_5$ matrix commutes with $J_\alpha$
in (\ref{spin-rotation}) and therefore $\gamma_5$ does not violate the spin symmetry. However such terms vanish
due to $\vslash\,H = H$ and the identity
\begin{eqnarray}
\frac{1+ \vslash}{2}  \,\gamma_5 \,\frac{1+ \vslash}{2}  = 0 \,.
\end{eqnarray}
Moreover the matrix $\gamma_\mu$ does not commute with $J_\alpha$. Nevertheless it sometimes leads to spin symmetric
terms. This is a consequence of the projection result
\begin{eqnarray}
\frac{1+ \vslash}{2} \,\gamma_\mu \,\frac{1+ \vslash}{2}  = v_\mu \,\frac{1+ \vslash}{2}  \,,
\end{eqnarray}
which, however, also shows that terms with $\gamma_\mu$ will not lead to further spin symmetric terms.

\begin{table}[t]
\setlength{\tabcolsep}{2mm}
\renewcommand{\arraystretch}{1.2}
\begin{center}
\begin{tabular}{l||c|c|l||c|c}\hline
$g_{0,[\bar3\bar3]}^{(S)} $&$  h_{2} $&$  -\frac{5}{6}\,g_2 - \frac{1}{8}\,g_+ $&$ M^{1/2}_{[\bar 3]}\,g_{0,[\bar3\bar3]}^{(V)} $&$ \frac12\,h_{4} $&$\,\frac43\,g_3 -\,3\,g_4 $\\
$g_{F,[\bar3\bar3]}^{(T)} $&$   0     $&$ 0 $&$ M^{1/2}_{[\bar 3]}\,g_{1,[\bar3\bar3]}^{(V)} $&$ - $&$ -$\\
$g_{D,[\bar3\bar3]}^{(S)} $&$  h_{1} $&$ -\,g_1 +\frac{1}{2}\,g_2 + \frac{3}{8}\,g_+ $&$ M^{1/2}_{[\bar 3]}\,g_{D,[\bar3\bar3]}^{(V)} $&$ \frac12\,h_{3} $&$ -3\,g_3 +\,\frac{9}{2}\,g_4 $\\

$g_{0,[66]}^{(S)} $&$  -h_{6}-\frac{1}{3}\,h_{13} $&$ 0$&$ M^{1/2}_{[6]}\, g_{0,[66]}^{(V)} $&$ \frac{1}{6}\,h_{13} - \frac12\,h_{9} $&$ 0$\\
$g_{1,[66]}^{(S)} $&$  -h_{7}-\frac{2}{3}\,h_{14} $&$  -\,g_2 + \frac{1}{12}\,g_+ $&$ M^{1/2}_{[6]}\,g_{1,[66]}^{(V)} $&$ \frac{1}{6}\,h_{14} - \frac14\,h_{10} $&$  -\,\frac{1}{2}\,g_4 $\\
$g_{D,[66]}^{(S)} $&$  -h_{5}-\frac{1}{3}\,h_{11} $&$ -\,g_1 -\frac{1}{2}\,g_2 + \frac{1}{8}\,g_+ $&$ M^{1/2}_{[6]}\,g_{D,[66]}^{(V)} $&$ \frac{1}{6}\,h_{11} - \frac12\,h_{8} $&$ -g_3 -\,\frac{1}{2}\,g_4 $\\

$g_{F,[\bar 36]}^{(T)} $&$ \frac{1}{\sqrt{3}}\,h_{15}  $&$ -\,\frac{1}{2\sqrt{3}}\,g_5 +\,\frac{1}{8\sqrt{3}}\,g_-  $&$ g_{F,[66]}^{(T)} $&$ \frac{1}{3}\,h_{12} $&$ \frac{1}{3}\,g_5 - \frac{1}{12}\,g_- $\\

$g_{1,[\bar 36]}^{(T)} $&$ \frac{1}{\sqrt{3}}\,h_{16}  $&$ \,\frac{1}{4\sqrt{3}}\,g_- $&$f_{1,[\bar 36]}^{(A)} $&$ h_{16} $&$ \frac{1}{4}\,g_- $\\

$ g_{1,[\bar 36]}^{(V)} $&$  0 $&$ 0 $&$f_{D,[\bar 36]}^{(A)} $&$ 0 $&$ 0 $\\
$ g_{D,[\bar 36]}^{(V)} $&$  0 $&$ 0 $&$f_{F,[\bar 36]}^{(A)} $&$ -\,2\,h_{15} $&$ g_5 - \frac{1}{4}\,g_- $\\
$ g_{D,[\bar 36]}^{(S)} $&$  0 $&$ 0 $&$ M^{1/2}_{[6]}\,h_{0,[66]}^{(V)} $&$ \frac{1}{2}\,h_{9} $&$ 0 $\\

$h_{0,[66]}^{(S)} $&$  h_{6} $&$ 0$&$ M^{1/2}_{[6]}\,h_{1,[66]}^{(V)} $&$ \frac14\,h_{10} $&$ \frac{1}{2}\,g_4 $\\
$h_{1,[66]}^{(S)} $&$  h_{13} $&$ 0 $&$ M^{1/2}_{[6]}\,h_{2,[66]}^{(V)} $&$ \frac{1}{2}\,h_8 $&$ \,g_3 +\frac{1}{2}\,g_4 $\\
$h_{2,[66]}^{(S)} $&$  h_{5} $&$ g_1 +\frac{1}{2}\,g_2 - \frac{1}{8}\,g_+ $&$ f_{0,[66]}^{(A)} $&$ \frac{1}{\sqrt 3}\,h_{13} $&$ 0 $\\
$h_{3,[66]}^{(S)} $&$  h_{11} $&$ 0 $&$ f_{1,[66]}^{(A)} $&$ \frac{1}{\sqrt 3}\,h_{14} $&$ \frac{1}{4\sqrt{3}}\,g_+ $\\
$h_{4,[66]}^{(S)} $&$  h_{7} $&$  g_2 - \frac{1}{4}\,g_+ $&$ f_{D,[66]}^{(A)} $&$ \frac{1}{\sqrt 3}\,h_{11} $&$ 0 $\\
$h_{5,[66]}^{(S)} $&$  h_{14} $&$ \frac{1}{4}\,g_+ $&$ f_{F,[66]}^{(A)} $&$ \frac{1}{\sqrt 3}\,h_{12} $&$ \frac{1}{\sqrt3}\,g_5 - \frac{1}{4\sqrt{3}}\,g_- $\\
$h_{F,[66]}^{(T)} $&$  -\,\frac{1}{2}\,h_{12} $&$ -\,\frac{1}{2}\,g_5 +\,\frac{1}{8}\,g_- $&$ g_{1,[\bar36]}^{(S)}  $&$ 0 $&$ 0 $\\ \hline

\end{tabular}
\caption{The symmetry conserving two-body counter terms as introduced in (\ref{def-L2}) are correlated by the heavy-quark symmetry
(\ref{def-LH-symmetric}) and the leading large-$N_c$ operators (\ref{def-AA-Nc}) with $g_\pm = g_6 \pm g_7$ and
$M_{[6]}^{1/2} = M_{[6]}^{3/2}$ in the heavy-quark mass limit. Note that only the combinations 
$g_{D,[\bar3\bar3]}^{(V)} +2\,g_{1,[\bar3\bar3]}^{(V)}$ and $g_{0,[\bar3\bar3]}^{(V)} -g_{1,[\bar3\bar3]}^{(V)}$ can be matched
at leading order in a non-relativistic expansion.}
\label{tab:two-body}
\end{center}
\end{table}

In a first step we reproduce previous results obtained in \cite{Yan:1992gz,Cho:1992gg,Jenkins:1996de}. There are two
spin-symmetric kinetic terms
\begin{eqnarray}
&& \mathscr{L}_{\rm kinetic}^{(H)} ={\rm tr}\,\bar H_{[\bar 3]}\,g_{\mu \nu}\,v^\mu\,i\,D^\nu \,H_{[\bar 3]}
-{\rm tr}\,\bar H_{[6]}^\mu\,g_{\mu \nu}\,(i\,D \cdot v ) \,H_{[6]}^\nu\,.
\label{kinetic lagrangian}
\end{eqnarray}
This implies that the two sextet masses $M^{1/2}_{[6]} $ and $M^{3/2}_{[6]} $ are degenerate in the
heavy-quark mass limit \cite{Yan:1992gz,Cho:1992gg,Jenkins:1996de}. According to (\ref{larg-Mc-scalaing}) the anti-triplet mass
$M^{1/2}_{[\bar 3]}$ remains as an independent parameter in that limit. We continue with
the constraint on the three-point vertices, for which the spin-symmetric interactions
\begin{eqnarray}
&& \mathscr{L}_{\rm 3-point}^{(H)} = i\,F_1\,\epsilon_{\mu\nu \alpha \beta}\,{\rm tr}\,\bar H_{[6]}^\mu\,v^\alpha\,i\,U^\beta\,H_{[6]}^\nu
+ F_2\,{\rm tr}\,\big( \bar H_{[6]}^\mu\,i\,U_\mu\,H^{\phantom{\mu}}_{[\bar 3]} + {\rm h.c.}\big) \,,
\label{3-pt lagrangian}
\end{eqnarray}
are parameterized by two coupling constants $F_1$ and $F_2$.
A matching of the Lagrangian (\ref{3-pt lagrangian}) with the corresponding terms in (\ref{def-L1})
leads to 4 sum rules
\begin{eqnarray}
&&F_1=  \textcolor{black}{-}\,H_{[66]} =   \sqrt 3\,C_{[66]} =  \textcolor{black}{-}\,\frac{3}{2}\,F_{[66]} \,,\qquad
 F_2 = C_{[\bar 36]} = \sqrt 3\,F_{[\bar 36]}   \,, \qquad
F_{[\bar 3\bar 3]} = 0\,,
\label{result-spin-symmetry-3-point}
\end{eqnarray}
with which we recover the results of \cite{Yan:1992gz,Cho:1992gg}.

We turn to the symmetry breaking counter terms introduced in (\ref{def-L2}, \ref{def-L4}),
for which we construct their spin-symmetric correspondence
\begin{eqnarray}
\mathscr{L}_{\chi}^{(H)} &=&  d_{1}\,{\rm tr}\,\big( \bar H_{[\bar 3]}\,H_{[\bar 3]}\Big)\,{\rm tr}\,\big(\chi_+\big)
+ d_{2}\,{\rm tr}\,\big( \bar H_{[\bar 3]}\,\chi_+\,H_{[\bar 3]}\big)
\nonumber\\
&\textcolor{black}{-}& d_{3}\,{\rm tr}\,\big( \bar H_{[6]}^\mu\,g_{\mu\nu}\,H_{[6]}^\nu\big)\,{\rm tr}\,\big(\chi_+\big)
\textcolor{black}{-} d_{4}\,{\rm tr}\,\big( \bar H_{[6]}^\mu\,g_{\mu\nu}\,\chi_+\,H_{[6]}^\nu\,\big)
\nonumber\\
&\textcolor{black}{-}& e_{1}\,{\rm tr}\,\big( \bar H_{[6]}^\mu\,g_{\mu\nu}\,H_{[6]}^\nu\big)\,{\rm tr}\,\big(\chi_+^2\big)
\textcolor{black}{-} e_{2}\,{\rm tr}\,\big( \bar H_{[6]}^\mu\,g_{\mu\nu}\,H_{[6]}^\nu \big)\,\big({\rm tr}\,\chi_+\big)^2
\nonumber\\
&\textcolor{black}{-}& e_{3}\,{\rm tr}\,\big( \bar H_{[6]}^\mu\,g_{\mu\nu}\,\chi_+\,H_{[6]}^\nu\big)\,{\rm tr}\,\big(\chi_+\big)
\textcolor{black}{-} e_{4}\,{\rm tr}\,\big( \bar H_{[6]}^\mu\,g_{\mu\nu}\,\chi_+^2\,H_{[6]}^\nu\big)
\nonumber\\
&\textcolor{black}{-}& e_{5}\,{\rm tr}\,\big( \bar H_{[6]}^\mu\,g_{\mu\nu}\,\chi_+\,H_{[6]}^\nu\,\chi^T_+\big)
+ e_{6}\,{\rm tr}\,\big( \bar H_{[\bar 3]}\,H_{[\bar 3]}\big)\,{\rm tr}\,\big(\chi_+^2\big)
\nonumber\\
&+& e_{7}\,{\rm tr}\,\big( \bar H_{[\bar 3]}\,H_{[\bar 3]}\big)\,\big({\rm tr}\,\chi_+\big)^2
+ e_{8}\,{\rm tr}\,\big( \bar H_{[\bar 3]}\,\chi_+\,H_{[\bar 3]}\,\big)\,{\rm tr}\,\big(\chi_+\big)
\nonumber\\
&+& e_9\,{\rm tr}\,\big( \bar H_{\bar3}\,\chi_+^2\,H_{\bar3}\big)
\,.
\label{def-LHchi}
\end{eqnarray}
There are 4 parameters, $d_n$, relevant at NLO and 9 parameters, $e_n$, at N$^3$LO in a computation of the baryon masses.
A matching of (\ref{def-LHchi}) with (\ref{def-L2}, \ref{def-L4}) determines 11 sum rules. There are 3 sum rules for the
NLO parameters
\begin{eqnarray}
&& b_{1,[66]} = d_{1,[66]} = d_3\,, \qquad  \qquad \quad
   b_{2,[66]} = d_{2,[66]} = d_4\,, \qquad \qquad  \quad
   b_{1,[\bar3 6]} = 0\,,
\nonumber\\
&& b_{1,[\bar 3\bar 3]} = d_1\,, \qquad  \qquad \qquad \qquad \;\;b_{2,[\bar 3\bar 3]} = d_2\,, \qquad \quad
\label{result-sumrule-d}
\end{eqnarray}
and 8 sum rules for the N$^3$LO parameters
\begin{eqnarray}
&& c_{n,[66]} = e_{n,[66]} = e_n \qquad {\rm for} \qquad n= 1,\cdots ,5 \,,\qquad
\nonumber\\
&& c_{n,[\bar 3 6]} = 0 \qquad  \qquad \qquad \, {\rm for} \qquad n= 1, \cdots ,3 \,,
\nonumber\\
&& c_{1,[\bar 3\bar 3]} = e_6\,, \qquad \qquad c_{2,[\bar 3\bar 3]} = e_7\,,\qquad \qquad
   c_{3,[\bar 3\bar 3]} = e_8\,, \qquad \qquad
   c_{4,[\bar 3\bar 3]} = e_9\,.
\label{result-sumrule-e}
\end{eqnarray}

We close this section with our analysis of the 4-point vertices in the chiral Lagrangian
as constructed in (\ref{def-L2}). Altogether there are 16 spin-symmetric terms
\begin{eqnarray}
\mathscr{L}^H_{4-point} &=& {\rm tr}\,\bar H_{[\bar 3]}\,\Big\{ h_{1}\,\big\{U_\mu,\,U_\nu\big\}\,H_{[\bar 3]}
+ h_{2}\,{\rm tr}\,\big(U_\mu\,U_\nu\big) \,H_{[\bar 3]}  \Big\}\,g^{\mu\nu}
\nonumber\\
&+& {\rm tr}\,\bar H_{[\bar 3]}\,\Big\{ h_{3}\,\big\{U_\mu,\,U_\nu\big\}\,H_{[\bar 3]}
+ h_{4}\,H_{[\bar 3]}\,{\rm tr}\,\big(U_\mu\,U_\nu\big)   \Big\}\,v^{\mu}\,v^\nu
\nonumber\\
&+& {\rm tr}\,\bar H_{[6]}^\alpha\,\Big\{ h_{5}\,\big\{U_\mu,\,U_\mu\big\}\,H_{[6]}^\beta
+ h_{6}\,H_{[6]}^\beta \,{\rm tr}\,\big(U_\mu\,U_\nu\big)
+ h_{7}\,U_\mu\,H^\beta_{[6]}\,U^T_\nu
\Big\}\,g_{\alpha \beta }\,g^{\mu\nu}
\nonumber\\
&+& {\rm tr}\,\bar H_{[6]}^\alpha\,\Big\{ h_{8}\,\big\{U_\mu,\,U_\nu\big\}\,H_{[6]}^\beta
+ h_{9}\,H_{[6]}^\beta\,{\rm tr}\,\big(U_\mu\,U_\nu\big)
+ h_{10}\,U_\mu\,H^\beta_{[6]}\,U^T_\nu
\Big\}\,g_{\alpha \beta }\,v^\mu \,v^\nu
\nonumber\\
&+& {\rm tr}\,\bar H_{[6]}^\alpha\,\Big\{ h_{11}\,\big\{U_\alpha,\,U_\beta\big\} + h_{12}\,\big[U_\alpha,\,U_\beta\big] + h_{13}\,{\rm tr}\,\big(U_\alpha\,U_\beta\big) \Big\}\,H_{[6]}^\beta
\nonumber\\
&+& {\rm tr}\,\bar H_{[6]}^\alpha\,\Big\{  h_{14}\,\Big( U_\mu\,H_{[6]}^\mu\,U^T_\alpha +  U_\alpha\,H_{[6]}^\mu\,U^T_\mu \Big) \Big\}
\nonumber\\
&+& {\rm tr}\,\bar H_{[6]}^\alpha\,\Big\{ h_{15}\,\big[U_\mu,\,U_\nu\big]\,H_{[\bar 3]}
+ h_{16}\,U_\mu\,H_{[\bar 3]}\,U^T_\nu
\Big\}\,i\,\epsilon_{\alpha \beta}^{\;\;\;\;\mu \nu }\,v^\beta + {\rm h.c.}\,,
\label{def-LH-symmetric}
\end{eqnarray}
where the conceivable two additional terms
\begin{eqnarray}
&& {\rm tr}\,\Big(\bar H_{[6]}^\alpha\,\big[U_\mu,\,U_\nu\big]\,H^\beta_{[6]}\Big) \,\epsilon_{\alpha \beta}^{\;\;\;\; \mu \nu }\,, \qquad \qquad
 {\rm tr}\, \Big(\bar H_{[6]}^\alpha\, U_\mu\,H^\beta_{[6]}\,U^T_\nu
\Big)\,\epsilon_{\alpha \beta}^{\;\;\;\; \mu \nu }\,,
\label{def-extra-terms}
\end{eqnarray}
violate parity conservation.
The result of a matching of the chiral Lagrangian (\ref{def-L2}) with (\ref{def-LH-symmetric})
is shown in Tab. \ref{tab:two-body}. This leads to $36-16 =20$  sum rules
\begin{eqnarray}
&& g_{0,[66]}^{(S)} = -h_{0,[66]}^{(S)} -\frac 13\,h_{1,[66]}^{(S)}\,,\qquad\quad
g_{1,[66]}^{(S)} = -h_{4,[66]}^{(S)} -\frac{2}{3}\,h_{5,[66]}^{(S)}\,,\qquad\quad
\nonumber\\
&& g_{D,[66]}^{(S)} = -h_{2,[66]}^{(S)} -\frac 13\,h_{3,[66]}^{(S)}\,,\qquad\quad
g_{1,[\bar36]}^{(S)} =  0\,,\qquad\quad
g_{D,[\bar36]}^{(S)} =  0\,,
\nonumber\\
&& M^{1/2}_{[6]}\,g_{0,[66]}^{(V)} = \frac{1}{6}\,h_{1,[66]}^{(S)} -M^{1/2}_{[6]}\,h_{0,[66]}^{(V)}\,,\qquad\quad
M^{1/2}_{[6]}\,g_{1,[66]}^{(V)} = \frac{1}{6}\,h_{5,[66]}^{(S)} -M^{1/2}_{[6]}\,h_{1,[66]}^{(V)}\,,\qquad\quad
\nonumber\\
&& M^{1/2}_{[6]}\,g_{D,[66]}^{(V)} =  \frac{1}{6}\,h_{3,[66]}^{(S)} -M^{1/2}_{[6]}\,h_{2,[66]}^{(V)}\,,\qquad\quad
g_{1,[\bar36]}^{(V)} =  0\,,\qquad\quad
g_{D,[\bar36]}^{(V)} =  0\,,\qquad\quad
\nonumber\\
&& f_{0,[66]}^{(A)} = \frac{1}{\sqrt3}\,h_{1,[66]}^{(S)}\,,\qquad\quad
f_{1,[66]}^{(A)} = \frac{1}{\sqrt3}\,h_{5,[66]}^{(S)}\,,\qquad\quad
f_{D,[66]}^{(A)} = \frac{1}{\sqrt3}\,h_{3,[66]}^{(S)}\,,\qquad\quad
\nonumber\\
&& g_{F,[66]}^{(T)} = \frac{1}{\sqrt3}\,f_{F,[66]}^{(A)}\,,\qquad\quad
h_{F,[66]}^{(T)} =  -\frac{3}{2}\,g^{(T)}_{F,66}\,,\qquad\quad
g_{F,[\bar36]}^{(T)} =  -\,\frac{1}{2\sqrt{3}}\,f_{F,[\bar36]}^{(A)}\,,\qquad\quad
\nonumber\\
&& g_{1,[\bar3\bar3]}^{(V)} = 
 g_{F,[\bar3\bar3]}^{(T)} = f_{D,[\bar36]}^{(A)} =  0\,,\qquad\quad
f_{1,[\bar36]}^{(A)} =  \,\sqrt{3} \,g_{1,[\bar36]}^{(T)}\,,
\label{HQS-baryon sumrules}
\end{eqnarray}
where  $M^{1/2}_{[6]}= M^{3/2}_{[6]}$ is charmed sextet baryon mass in the heavy-quark mass limit.

Altogether the $3+6+7+36+17= 69$ distinct low-energy constants were correlated by $1 + 4 + 3 + 20 + 8=36$ sum rules that
follow from the heavy-quark spin symmetry. That leaves 33 unknown parameters.
It remains to work out additional sum-rules that arise in the large-$N_c$ limit of QCD.

\newpage

\section{Large-$N_c$ operator analysis }

In this section we further correlate the parameters of the effective interaction introduced in
Section \ref{section:chiral-lagrangian}. We follow the works of Luty and March-Russell \cite{Luty1994} and
of Dashen, Jenkins and Manohar \cite{Dashen1994,Jenkins:1996de}. These works introduced a formalism for a
systematic expansion of baryon matrix elements of QCD quark operators in powers of $1/N_c$. In previous works
applications of this formalism to correlation functions involving a product of two axial-vector or vector quark
currents were worked out \cite{Lutz:2010se,Lutz:2011fe}.

The $1/N_c$ expansion of an $l$-body QCD operator takes the generic form
\begin{eqnarray}
{\mathcal O}^{(l)}_{QCD} = N_c^l \sum_{m=0}^{N_l} \,\sum_{n=0}^{N_h}\,c^{(m+n)}\,\frac{1}{N_c^{m+n}}\,
{\mathcal O}_l^{(m)}\,{\mathcal O}_h^{(n)}\,,
\label{def-largeN-expansion}
\end{eqnarray}
where ${\mathcal O}_l^{(m)}$  and ${\mathcal O}_h^{(n)}$ are suitable $m$- and $n$-body operators
composed of light and heavy-quark field components respectively. In (\ref{def-largeN-expansion}) it is assumed
that the operators act on baryon states containing $N_l$ and $N_h$ light and heavy quarks with $N_c = N_l + N_h$.

It was shown in  \cite{Luty1994,Dashen1994,Jenkins:1996de,Lutz:2010se} that the operator
expansion (\ref{def-largeN-expansion}) can be performed most efficiently in terms of a complete set of
static and color-neutral one-body operators that act on effective baryon states rather than the physical states.
While the operator on the left-hand side of (\ref{def-largeN-expansion}) is to be evaluated in the physical states,
the right-hand side was rewritten such that only effective static baryon states occur. In our case the physical and
effective baryon states
\begin{eqnarray}
\ket{p,\, ij_\pm ,\,S,\,\chi }\,,  \qquad \qquad \roundket{ij_\pm,\,S,\, \chi}\,,
\label{def-states}
\end{eqnarray}
are specified by the momentum $p$ and the flavor indices $i,j,k=1,2,3$. The spin $S$ and the spin-polarization
are $\chi = 1,2$ for the spin one-half ($S=1/2$) and $\chi =1,\cdots ,4$ for the spin three-half states $(S=3/2)$. The
flavour sextet and the anti-triplet are discriminated by their symmetric (index $+$) and anti-symmetric (index $-$)
behaviour under the exchange of $i \leftrightarrow j$. At leading order in the $1/N_c$ expansion all considered baryon
states are mass degenerate. It is important to note that unlike the physical baryon states,
the effective baryon states do not depend on the momentum $p$. All dynamical information is moved into the
coefficient functions.

The effective baryon states $\roundket{ij_\pm, \chi}$ were shown to have a mean-field structure that can be
generated in terms of effective quark operators $q$ and $Q$ for the light and heavy species respectively. A corresponding
complete set of color-neutral one-body operators was constructed in terms of the very same static quark operators
\begin{eqnarray}
&& \quarknumberoperator = q^\dagger ( \one  \otimes \one  \otimes \one )\,q \,, \qquad  \qquad \;\;\,
J_i = q^\dagger \Big(\frac{\sigma_i }{2} \otimes \one \otimes \one \Big)\, q \,,
\nonumber\\
&& T^a = q^\dagger \Big(\one \otimes \frac{\lambda_a}{2} \otimes \one \Big)\, q\, ,\qquad \quad \;\;
G^a_i = q^\dagger \Big( \frac{\sigma_i}{2} \otimes \frac{\lambda_a}{2} \otimes \one \Big)\, q\,,
\nonumber\\
&& \quarknumberoperator_{\!Q} = Q^\dagger ( \one  \otimes \one )\,Q \,, \qquad  \qquad \;\;\;\;\;\,
J_Q^{(i)} = Q^\dagger \Big(\frac{\sigma^{(i)} }{2} \otimes \one \Big)\, Q \,,
\label{def:one-body-operators}
\end{eqnarray}
with the quark operators $q=(u,d,s)^t$ and $Q= c$ of the up, down, strange and charm quarks. With $\lambda_a$ we
denote the Gell-Mann matrices supplemented with a singlet matrix $\lambda_0 = \sqrt{2/3}\,\one $.
Here we use a redundant notation with
\begin{eqnarray}
T^0 = \sqrt{\frac{1}{6}}\,\quarknumberoperator \,, \qquad \qquad G^0_i = \sqrt{\frac{1}{6}}\,J_i \,,
\label{def-redundant-operator}
\end{eqnarray}
which will turn useful when analyzing matrix elements of scalar currents.

All what is needed in any practical application of the $1/N_c$ expansion is the action of any of the
one-body operators introduced in (\ref{def:one-body-operators}) on the effective mean-filed type baryon states
$\roundket{ij_\pm, \chi}$. In fact it suffices to provide results at the physical value $N_c =3$, for which we generated
the following complete list
\allowdisplaybreaks
\begin{eqnarray}
&& \quarknumberoperator \,|\,ij_{\pm},\,{\textstyle{1\over 2}},\,\chi\,\big)
= 2\,|\,ij_{\pm },\,{\textstyle{1\over 2}},\,\chi\,\big)\,, \qquad \qquad \qquad \;\;\,
\quarknumberoperator \,|\,ij_{+},\,{\textstyle{3\over 2}},\,\chi\,\big)
= 2\,|\,ij_{+ },\,{\textstyle{3\over 2}},\,\chi\,\big)\,,
\nonumber\\
&& \quarknumberoperator_{\!Q} \,|\,ij_{\pm},\,{\textstyle{1\over 2}},\,\chi\,\big)
= |\,ij_{\pm },\,{\textstyle{1\over 2}},\,\chi\,\big)\,, \qquad \qquad \qquad \;\;\;
\quarknumberoperator_{\!Q} \,|\,ij_{+},\,{\textstyle{3\over 2}},\,\chi\,\big)
= |\,ij_{+ },\,{\textstyle{3\over 2}},\,\chi\,\big) \,,
\nonumber\\
&& T^a\,|\,ij_{\pm},\,{\textstyle{1\over 2}},\,\chi\,\big)
= \Lambda_{(ij)_\pm}^{(a),\,(mn)_\pm }\,|\,mn_\pm,\,{\textstyle{1\over 2}},\,\chi\,\big) \,, \qquad
T^a\,|\,ij_{+},\,{\textstyle{3\over 2}},\,\chi\,\big)
= \Lambda_{(ij)_+}^{(a),\,(mn)_+ }\,|\,mn_+,\,{\textstyle{3\over 2}},\,\chi\,\big) \,,
\nonumber\\
&& J_Q^{(k)}\,|\,ij_\pm,\,{\textstyle{1\over 2}},\,\chi\,\big)
= \frac 12\,\sigma^{(k)}_{\bar\chi\chi}\,|\,ij_\pm,\,{\textstyle{1\over 2}},\,\bar\chi\,\big)\,,\qquad  \quad \;\;\,
\nonumber\\
&& J_Q^{(k)}\,|\,ij_+,\,{\textstyle{3\over 2}},\,\chi\,\big)
= \frac 12\,\big(\vec S\,\sigma^{(k)}\,\vec S^{\,\dagger}\,\big)_{\bar\chi\chi}\,|\,ij_+,\,{\textstyle{3\over 2}},\,\bar\chi\,\big)
+\textcolor{black}{\frac{ 1}{ \sqrt{3}}}\, S^{ (k) \dagger}_{\bar\chi\chi}\,|\,ij_{+},\,{\textstyle{1\over 2}},\,\bar \chi\,\big)\,,
\nonumber\\ \nonumber\\
&& J^k\,|\,ij_+,\,{\textstyle{1\over 2}},\,\chi\,\big)
= \frac 23\,\sigma_{\bar\chi\chi}^{(k)}\,|\,ij_+,\,{\textstyle{1\over 2}},\,\bar\chi\,\big)
-\,\frac{1}{\sqrt 3}\,\,S_{\bar\chi\chi}^{(k)}|\,ij_+,\,{\textstyle{3\over 2}},\,\bar\chi\,\big)\,,
\nonumber\\
&& G_k^a\,|\,ij_+,\,{\textstyle{1\over 2}},\,\chi\,\big) = \frac 13\,\Lambda_{(ij)_+}^{(a),\,(mn)_+}\,\sigma_{\bar\chi\chi}^{(k)}\,|\,mn_+,\,{\textstyle{1\over 2}},\,\bar\chi\,\big)
-\,\frac{1}{2\sqrt 3}\,\Lambda_{(ij)_+}^{(a),\,(mn)_+}\,S_{\bar\chi\chi}^{(k)}\,|\,mn_+,\,{\textstyle{3\over 2}},\,\bar\chi\,\big)
\nonumber\\
&& \qquad \qquad  -\,\frac{1}{2\,\sqrt{3}}\,\Lambda_{(ij)_+}^{(a),\,(mn)_-}\,\sigma_{\bar\chi\chi}^{(k)}\,
|\,mn_-,\,{\textstyle{1\over 2}},\,\bar\chi\,\big)\,,
\nonumber\\ \nonumber\\
&& J^k\,|\,ij_+,\,{\textstyle{3\over 2}},\,\chi\,\big)
= \big(\vec S\,\sigma^{(k)}\,\vec S^{\,\dagger}\,\big)_{\bar\chi\chi}\,|\,ij_+,\,{\textstyle{3\over 2}},\,\bar\chi\,\big)
-\,\frac{1}{\sqrt{3}}\,S_{\bar\chi\chi}^{(k)\,\dagger}\,|\,ij_+,\,{\textstyle{1\over 2}},\,\bar\chi\,\big)
\nonumber\\
&& G_k^a\,|\,ij_+,\,{\textstyle{3\over 2}},\,\chi\,\big)
= \frac 12\,\Lambda_{(ij)_+}^{(a),\,(mn)_+}\,(\vec S\,\sigma^{(k)}\,\vec S^\dagger\,)_{\bar\chi\chi}\,|\,mn_+,\,{\textstyle{3\over 2}},\,\bar\chi\,\big)
-\,\frac{1}{2\sqrt 3}\,\Lambda_{(ij)_+}^{(a),\,(mn)_+}\,S_{\bar\chi\chi}^{(k)\dagger}\,|\,mn_+,\,{\textstyle{1\over 2}},\,\bar\chi\,\big)
\nonumber\\
&&\qquad \qquad -\,\frac{1}{2}\,\Lambda_{(ij)_+}^{(a),\,(mn)_-}\,S_{\bar\chi\chi}^{(k)\dagger}
\,|\,mn_-,\,{\textstyle{1\over 2}},\,\bar\chi\,\big)\,,
\nonumber\\ \nonumber\\
&& J^k\,|\,ij_-,\,{\textstyle{1\over 2}},\,\chi\,\big)
= 0 \,,
\nonumber\\
&& G_k^a\,|\,ij_-,\,{\textstyle{1\over 2}},\,\chi\,\big)
= -\,\frac{1}{2\sqrt 3}\,\Lambda_{(ij)_-}^{(a),\,(mn)_+}\,\sigma_{\bar\chi\chi}^{(k)}\,|\,mn_+,\,{\textstyle{1\over 2}},\,\bar\chi\,\big)
-\,\frac 12\,\Lambda_{(ij)_-}^{(a),\,(mn)_+}\,S_{\bar\chi\chi}^{(k)}\,|\,mn_+,\,{\textstyle{3\over 2}},\,\bar\chi\,\big)\,,
\label{result-state-action}
\end{eqnarray}
in terms of the notation introduced already in (\ref{def:spin-transition-matrices}, \ref{def-one-pm}).
Note that all results (\ref{result-state-action}) are
valid also for the singlet components with $a=0$ if the convention $\lambda_0 = \sqrt{2/3} \,\one$ is used in
(\ref{def:spin-transition-matrices}).

In the sum of (\ref{def-largeN-expansion}) there are infinitely many terms one may write down.
Terms that break the heavy-quark spin symmetry are exclusively caused by the heavy-spin operator
\begin{eqnarray}
J^i_Q \sim \frac{1}{M_Q} \,,
\label{def-spin-violation}
\end{eqnarray}
with the heavy-quark mass $M_Q$. In contrast the counting of $N_c$ factors is intricate since there is
a subtle balance of suppression and enhancement effects.
An $r$-body operator consisting of the $r$ products of any of the spin and flavor operators receives the
suppression factor $N_c^{-r}$. This is counteracted by enhancement factors for the
flavor and spin-flavor operators $T^a$ and $G^a_i$ that are
produced by taking baryon matrix elements at $N_c \neq 3$.
Altogether this leads to the effective scaling laws \cite{Dashen1994,Jenkins:1996de}
\begin{eqnarray}
J_i \sim \frac{1}{N_c} \,, \qquad \quad T^a \sim N^0_c \,, \qquad \quad G^a_i \sim N^0_c \,.
\label{effective-counting}
\end{eqnarray}
According to (\ref{effective-counting}) there is an infinite number of terms contributing at a given order in the
the $1/N_c$ expansion. Taking higher products of flavor and spin-flavor operators does not reduce the $N_c$
scaling power. A systematic $1/N_c$ expansion is made possible by a set of operator
identities \cite{Dashen1994,Jenkins:1996de,Lutz:2010se}, that allows a systematic summation of the infinite number
of relevant terms.

The reduction algorithm of \cite{Dashen1994,Jenkins:1996de,Lutz:2010se} relies on an analysis of the product
of two one-body-operators. First, antisymmetric products are considered. They can always be
reduced to a one-body operator by using the Lie-algebra relations
\begin{eqnarray}
&&\big[J_i,\, J_j \big]=i\,\varepsilon_{ijk}\, J_k, \qquad\;\; \big[T^a,\, T^b\big]=i\,f_{abc}\, T^c, \qquad
\big[J^i,\, T^a\big]=0\,,
\nonumber \\
& &\big[J_i,\, G^{a}_j\big]=i\,\varepsilon_{ijk}\, G^{a}_k\,,\qquad
\big[T^a,\, G^{b}_i\big]=i\,f_{abc}\, G^c_{i}\,,
\nonumber \\
&&\big[G^{a}_i,\, G^{b}_j\big]=\frac{i}{4}\,\delta_{ij}\, f_{abc}\, T^c + \frac{i}{6}\, \delta_{ab}\,
\varepsilon_{ijk}\, J_k + \frac{i}{2}\, \varepsilon_{ijk}\, d_{abc}\, G_{k}^c\,,
\nonumber\\
&& \big[J^{(i)}_Q,\, J^{(j)}_Q \big]=i\,\varepsilon_{ijk}\, J^{(k)}_Q, \qquad\;
\big[J^{(i)}_Q,\,J_i \big] = \big[J^{(i)}_Q,\,T^a \big] = \big[J^{(i)}_Q,\,G^a_i \big] =0 \,,
\label{def:SU(6)-Lie-algebra}
\end{eqnarray}
which may be verified by an explicit computation as a consequence of the canonical
anti-commutator relations of the static quark operators.

It remains to consider the symmetric products of two one-body operators, which were studied in great depth in
\cite{Dashen1994,Lutz:2010se}. A simple example of such identities is the anti-commutator of two heavy-spin operators
\begin{eqnarray}
\textcolor{black}{\{J^i_Q, \,J^i_Q\}} = \frac{3}{2}\,\quarknumberoperator_{\!Q} \,.
\label{def-anti}
\end{eqnarray}
While all Lie-algebra identities (\ref{def:SU(6)-Lie-algebra}) hold as such, the identity (\ref{def-anti}) holds only
if matrix elements in the charmed baryon states introduced in (\ref{def-states}) are taken.
As was observed by Jenkins \cite{Jenkins:1996de} the relations obtained
in \cite{Dashen1994} can be adapted to the present case of matrix elements in charmed baryons.
Formally they follow from the results \cite{Dashen1994,Lutz:2010se} upon the replacement $N_c \to N_l$.
The complete set of operator identities reads
\begin{eqnarray}
\delta_{ab}\,\{T^a, \,T^b\} &=& \frac 16\, (N_l+6)\, \quarknumberoperator + \{J_k,\, J_k\}\,,
\nonumber \\
d_{abc}\, \{T^a, T^b\} &=& -\frac 13\, (N_l+3)\, T^c + 2\,\{ J^i , G^{ic} \}\,,
\label{result:reduction-identities:1}
\end{eqnarray}
and
\begin{eqnarray}
\delta_{ab}\,\{T^a,\, G^{b}_i\} &=& \frac 23\,(N_l+3)\, J_i \,,
\nonumber \\
d_{abc}\, \{T^a, G^{b}_i\} &=& \frac 13\, (N_l+3)\, G^c_i + \frac 16 \{J_i, \,T^c\}\,,
\nonumber \\
f_{abc} \,\{T^a, G^{b}_i\} &=& \varepsilon_{ijk}\, \{J_j, \,G^c_{k}\}\,,
\label{result:reduction-identities:2}
\end{eqnarray}
and
\begin{eqnarray}
\delta_{ab}\,\{G^{a}_i, \,G^{b}_{j}\} &=& \frac{1}{8}\,\delta_{ij}\, \Big( (N_l+6)\, \quarknumberoperator
- 2\, \{J_k,\, J_k\} \Big) + \frac 13 \, \{J_i, J_j\}\,,
\nonumber \\
d_{abc}\,\{G^{a}_i, \,G^{b}_j\} &=& \frac 13\,\delta_{ij}\, \Big( \frac 43 \,(N_l+3)\, T^c - \frac 32 \,\{J_k,\, G^c_k\} \Big)
\nonumber \\
&+&\frac 16 \,\Big( \{J_i, \,G^c_j \} + \{J_j,\, G^c_i\} \Big)\,,
\nonumber \\
 \{G^{a}_k,\, G^{b}_k\} &=& \frac{1}{24}\, \delta_{ab}\,\Big( (N_l+6)\, \quarknumberoperator - 2\, \{J_k, J_k\} \Big)
\nonumber \\
&+& \frac 12 \,d_{abc}\,\Big( (N_l+3)\, T^c - 2\, \{J_k, \,G^c_k\} \Big) + \frac 14 \,\{T^a, T^b\}\,,
\nonumber \\
 \varepsilon_{ijk}\,\{G^{a}_j,\, G^{b}_k\}
&=& \frac 12\,  f_{abc}\,\Big(- (N_l+3)\, G^c_{i}+ \frac{1}{6} \,\{J_i,\, T^c\} \Big)
\nonumber \\
&+& \frac 12\, \Big(f_{acg}\,d_{bch}-f_{bcg}\,d_{ach}\Big)\, \{T^g,\, G_{i}^h\} \,,
\label{result:reduction-identities:3}
\end{eqnarray}
which hold in matrix elements of the baryon ground-state tower. All identities are confirmed by
taking matrix elements in the charmed baryon states  with $N_l = 2$. Unlike in (\ref{result-state-action}) all flavour indices
in (\ref{result:reduction-identities:1}-\ref{result:reduction-identities:3}) exclude the singlet component.
Detailed expressions are collected in the Appendix.

Owing to the operator identities the general two reduction rules of \cite{Dashen1994} are recovered:
\begin{itemize}
\item All operator products in which two flavor indices are contracted using $\delta_{ab}$,
$f_{abc}$ or $d_{abc}$ or two spin indices on $G$'s are contracted using $\delta_{ij}$ or $\varepsilon_{ijk}$
can be eliminated.
\item All operator products in which two flavor indices are contracted using symmetric or antisymmetric
combinations of two different $d$ and/or $f$ symbols can be eliminated. The only exception to this rule is
the antisymmetric
combination $f_{acg}\,d_{bch}-f_{bcg}\,d_{ach}$.
\end{itemize}
As a consequence the infinite tower of spin-flavor operators truncates at any given order in the $1/N_c$ expansion.
We can now turn to the $1/N_c$ expansion of the baryon matrix elements of the QCD's axial-vector and scalar currents.
In application of the operator reduction rules, the baryon matrix elements of time-ordered products of the current
operators are expanded in powers of the effective one-body operators according to the counting rule (\ref{def-spin-violation}, \ref{effective-counting})
supplemented by the reduction rules (\ref{def:SU(6)-Lie-algebra}, \ref{def-anti}, \ref{result:reduction-identities:1}, \ref{result:reduction-identities:2}, \ref{result:reduction-identities:3}).
In contrast to Jenkins \cite{Jenkins:1996de} we consider the ratio $N_l/N_c= 1-1/N_c$ not as a suppression factor. The
strength of the spin-symmetry breaking terms we estimate with $1/M_Q \sim 1/N_c$. In the course of the construction of
the various structures, parity and time-reversal transformation properties are taken into account.

A first simple application of the operator reduction rules follows with matrix elements of the
axial-vector current (\ref{def-amu}). Keeping terms that are leading and subleading in the $1/N_c$ expansion
there are three relevant operators only
\begin{eqnarray}
&& \bra{\,\bar p,\,mn_\pm ,\,\bar S,\,\bar\chi\,}\,A_i^{(a)}(0)\,
\ket{\,p\,,kl_\pm, \, S,\,\chi \,}
=  \roundbra{\,mn_\pm,\,\bar S,\,\bar\chi \,} \,f_1\,G^{a}_i  + f_2\,\big\{ J_i,\, T^a \big \}
\nonumber\\
&& \qquad \qquad \qquad +\, f_3\,\big\{ J_{Q}^i,\, T^a \big\}
\roundket{\,kl_\pm,\,S,\,\chi\,} + \cdots \,.
\label{def-Amu-Nc}
\end{eqnarray}
The two spin-symmetric structures are parameterized by $f_{1,2}$ and the one spin-symmetry breaking term with
$f_3$. Since the last term is suppressed in the heavy-quark mass limit we expect the third structure to be of
minor importance only. We would include it at a level where 3-body operators that are further suppressed in the $1/N_c$
expansion are considered. It is instructive to match the three coupling constants to the 6 parameters
introduced in (\ref{def-L1}). A comparison of (\ref{axial-current element non-relativistic}) with
matrix elements of (\ref{def-Amu-Nc}) as provided in the Appendix leads to the following identification
\begin{eqnarray}
&&F_{[66]} = \textcolor{black}{\frac 23}\,f_1 + \textcolor{black}{\frac 83}\,f_2 + \textcolor{black}{2}\,f_3\,, \qquad
F_{[\bar 3\bar 3]} = \textcolor{black}{2}\,f_3\,, \qquad
 F_{[\bar 36]} = \textcolor{black}{-\frac{1}{\sqrt 3}}\,f_1\,,
\nonumber\\
&& C_{[66]} = \textcolor{black}{-\frac{1}{\sqrt 3} \,f_1 - \frac{4}{\sqrt 3}\,f_2 +\frac{2}{\sqrt 3}\,f_3 }\,, \qquad
C_{[\bar 3 6]} = \textcolor{black}{-\,f_1 }\,,\qquad
 H_{[66]} = \textcolor{black}{ f_1 + 4\,f_2 + 2\,f_3}\,,
\nonumber\\
&& C_{[66]} = \frac{1}{\sqrt 3}\, \Big( 2\,F_{[\bar 3\bar 3]}- H_{[66]}\Big) \,, \qquad  F_{[\bar 3\bar 3]} = 3\, F_{[66]}- 2\,H_{[66]}\,,
\qquad F_{[\bar 36]} = \frac{1}{\sqrt 3}\,C_{[\bar 36]}\,,
\label{result-large-Nc-axial}
\end{eqnarray}
It is reassuring that the spin-symmetry sum rules (\ref{result-spin-symmetry-3-point}) are recovered only
and only with $f_3=0$, which is consistent with the scaling ansatz (\ref{def-spin-violation}).

We turn to matrix elements of the scalar current (\ref{def-amu}, \ref{scalar-current element non-relativistic}).
At N$^2$LO in the $1/N_c$ expansion there are 5 operators to be considered
\begin{eqnarray}
&& \bra{\,\bar p,\,mn_\pm ,\,\bar S,\,\bar\chi\,}\,S^{(a)}(0)\,
\ket{\,p\,,kl_\pm, \, S,\,\chi \,}
=  \roundbra{\,mn_\pm,\,\bar S,\,\bar\chi \,} \,\delta_{a0}\,\big( b_1\, \quarknumberoperator  + b_2\,J^2 \big)
\nonumber\\
&& \qquad \qquad \qquad+ \,b_3 \,T^a  + b_4 \big\{J^i\,,G^a_i\big\}+  b_5\, \big\{J^i_Q\,,G^a_i\big\}
\roundket{\,kl_\pm,\,S,\,\chi\,} + \cdots  \,.
\label{def-S-Nc}
\end{eqnarray}
Here we consider the spin-symmetry breaking operator $\big\{J^i_Q\,,G^a_i\big\}$ since our counting suggests
that it is as important as the $J^2$ operator. The flavour index $a$ in (\ref{def-S-Nc}) may take the singlet value
$a=0$ with the singlet operators assumed in the notation (\ref{def-redundant-operator}).
The matrix elements of the operators in the baryon states are readily
looked up in the Appendix. A comparison with the tree-level expressions (\ref{scalar-current element non-relativistic})
generates the identifications
\begin{eqnarray}
&& b_{1,[66]} =  \textcolor{black}{-\,2\sqrt{\frac23}}\,\big(b_1 + \textcolor{black}{b_2} \big)\,,\qquad \qquad
b_{2,[66]} = \textcolor{black}{-\,2}\,\big( b_3 + 2\,b_4 + b_5 \big) \,,
\nonumber\\
&& d_{1,[66]} =  \textcolor{black}{-\,2\sqrt{\frac23}}\,\big(b_1 + \textcolor{black}{b_2} \big)\,, \qquad \qquad
d_{2,[66]} = \textcolor{black}{-\,2}\,\big( b_3 + 2\,b_4 + \textcolor{black}{{\textstyle{1 \over 2}}}\,b_5 \big)\,,
\nonumber\\
&& b_{1,[\bar3\bar3]} = \textcolor{black}{-\,2\sqrt{\frac23}}\,b_1 \,,\qquad \quad
b_{2,[\bar3\bar3]} = \textcolor{black}{-\,2}\,b_3 \,, \qquad \quad  b_{1,[\bar3 6]} = \textcolor{black}{\sqrt{3}}\,b_5\,.
\label{res-b-id}
\end{eqnarray}
The result (\ref{res-b-id}) implies 2 sum rules, which we already derived in (\ref{result-sumrule-d}) based on
the heavy-quark spin symmetry. The third sum rule in (\ref{result-sumrule-d}) is recovered with and only with  $b_5 = 0$.
Again this is consistent with the scaling ansatz (\ref{def-spin-violation}). Our results (\ref{res-b-id}) are related to the
previous analysis of Jenkins \cite{Jenkins:1996de}, that considered charmed baryon masses rather than matrix elements of
scalar currents in the $1/N_c$ expansion. Note that the counter terms in (\ref{res-b-id}) contribute
to the baryon self energies at tree-level already. Therefore a comparison with the operator analysis suggested in
\cite{Jenkins:1996de} is possible. Our analysis differs from the one in \cite{Jenkins:1996de} to the extent that our
expansion is not relying on an additional expansion in a quark-mass difference.

We continue with a derivation of large-$N_c$ sum rules for the chiral-symmetry breaking low-energy constants
introduced in (\ref{def-L4}). They contribute to the time-ordered product
of two scalar currents as evaluated in the baryon states (\ref{non-relativistic-correlator-SS}).
At NLO in the $1/N_c$ expansion we find the relevance of 5 operators
\begin{eqnarray}
&& \bra{\,\bar p,\,mn_\pm,\, \bar S,\,\bar\chi\,}\,S^{ab}(q)\,
\ket{p,\,kl_\pm,\, S,\,\chi\,} =
\roundbra{\,mn_\pm,\,\bar S,\,\bar\chi \,} \,{\mathcal O}^{ab} \,
\roundket{\,kl_\pm,\,S,\,\chi\,}\,,
\nonumber\\
&& \quad \,\textcolor{black}{+}\,{\mathcal O}^{ab} = \big( c_1\, \delta_{a0}\,\delta_{b0} + c_2\,\delta_{ab} \big)\, \quarknumberoperator
 + c_3 \,\big( T^a\, \delta_{b0}+ \delta_{a0}\,T^b \big)
\nonumber\\
&& \qquad \quad +\, c_4\, d_{abc}\,  T^c + c_5 \,\big\{T^a\,,T^b \big\}    \,,
\label{def-SS-Nc}
\end{eqnarray}
where at this order no spin-symmetry breaking operator has to be considered. The sum in (\ref{def-SS-Nc})
starts at $c=1,... , 8$. The matrix elements of the operators in (\ref{def-SS-Nc}) are given
in the Appendix. A comparison with the tree-level expressions (\ref{non-relativistic-correlator-SS})
implies the matching conditions
\begin{eqnarray}
&& c_{1,[\bar 3\bar 3]} = 2\,c_2+ {\textstyle{ 2 \over  3}}\,c_5   \,, \qquad \qquad
c_{2,[\bar 3\bar 3]} = {\textstyle{ 2 \over  3}}\,c_1 \,, \qquad \qquad
c_{3,[\bar 3\bar 3]} = 2\sqrt{{\textstyle{ 2 \over  3}}}\,c_3 \,,
\nonumber\\
&& c_{4,[\bar 3\bar 3]} = c_4 + c_5 \,,\qquad \qquad \quad \;\; c_{1,[\bar 36]} =
c_{2,[\bar 36]} = c_{3,[\bar 36]} = 0\,,
\nonumber\\
&& c_{1,[66]} = e_{1,[66]} =2\,c_2  +{\textstyle{ 2 \over  3}}\,c_5  \,, \qquad \qquad
c_{2,[66]} =e_{2,[66]} = {\textstyle{ 2 \over  3}}\,c_1 \,,
\nonumber\\
&& c_{3,[66]} = e_{3,[66]} =2\sqrt{{\textstyle{ 2 \over  3}}}\,c_3  \,, \qquad \qquad \quad \;
c_{4,[66]} = e_{4,[66]} =c_4 + c_5   \,,
\nonumber\\
&& c_{5,[66]} = e_{5,[66]} =  2\,c_5 - 6\,c_2  - 2\,c_4  \,,
\label{res-Nc-c}
\end{eqnarray}
From (\ref{res-Nc-c}) one can deduce $17 - 5 = 12$ sum-rules. This is to be compared with the $17 - 9 = 8$ sum rules
in (\ref{result-sumrule-e}) that follow from the heavy-quark spin symmetry only. Here large-$N_c$ QCD shows its predictive
power in generating 4 additional sum rules. We recover the 8 sum rules collected in (\ref{result-sumrule-e}) together with
the four additional sum rules
\begin{eqnarray}
&& c_{n,[\bar 3\bar 3]} = c_{n,[66]}  \qquad  {\rm for } \quad n = 1, \cdots , 4 \,\,.
\end{eqnarray}

We close this section with a study of the time-ordered product of two axial-vector currents.
The leading order operator expansion was already worked out in \cite{Lutz:2010se}. Here  we need to
consider matrix elements in charmed baryons and derive the implication for the chiral two-body interactions
introduced (\ref{def-L2}). According to \cite{Lutz:2010se} there are 7 distinct operators
\begin{eqnarray}
&& \bra{\,\bar p,\,mn_\pm,\, \bar S,\,\bar\chi\,}\,A_{ij}^{ab}(q)\,
\ket{p,\,kl_\pm,\, S,\,\chi\,} =
\roundbra{\,mn_\pm,\,\bar S,\,\bar\chi \,} \,{\mathcal O}^{ab}_{ij} \,
\roundket{\,kl_\pm,\,S,\,\chi\,}\,,
\nonumber\\
&& \quad \,{\mathcal O}^{ab}_{ij} = - \delta_{ij}\, \Big( g_1\,\big( {\textstyle{1\over 3}}\,\delta_{ab}\,\quarknumberoperator+
d_{abc}\,T^c \big) + {\textstyle{1\over 2}}\,g_2\,\big\{T^a,\,T^b\big\} \Big)
\nonumber\\
&& \qquad \quad+\, \frac{(\bar p+p)_i \,(\bar p+p)_j}{4\,M^2} \,\Big(
\,g_3\,\big( {\textstyle{1\over 3}}\,\delta_{ab}\,\quarknumberoperator +
d_{abc}\,T^c \big) + {\textstyle{1\over 2}}\,g_4\,\big\{T^a,\,T^b \big\}\,\Big)
\nonumber\\
&& \qquad \quad+\, \epsilon_{ijk}\, f_{abc}\, \,g_5\,G^c_k\,
+  {\textstyle{1\over 2}}\,g_6\,\big\{G^{a}_i,\,G^{b}_j\big\}
+ {\textstyle{1\over 2}}\,g_7\,\big\{G^{a}_j,\,G^{b}_i \big\}+ \cdots \,,
\label{def-AA-Nc}
\end{eqnarray}
where we focus on the space components of the correlation function. In (\ref{def-AA-Nc})
we have $q= \bar p-p$ and $a,b =1, \cdots ,8$. In addition, we consider terms only that arise in the small-momentum
expansion and that are required for the desired matching with (\ref{non-relativistic-correlator}).
The dots in (\ref{def-AA-Nc}) represent additional terms that are further suppressed in $1/N_c$ or for small
3-momenta $p$ and $\bar p$. Here we also assumed a degenerate baryon mass $M$ for the baryon states as they arise in
the large-$N_c$ limit. An application of the results of our Appendix leads to the matching result already included in
Tab. \ref{tab:two-body}, where they are compared with the consequences of the spin-symmetric interactions introduced
in (\ref{def-LH-symmetric}). From the operator analysis (\ref{def-AA-Nc}), we obtain 29 sum rules
\begin{eqnarray}
&& g_{D,[\bar 3\bar 3]}^{(S)} = -\,h_{2,[66]}^{(S)}+ h_{4,[66]}^{(S)} +2\,h_{5,[66]}^{(S)}\,,\qquad\quad 
g_{0,[\bar3\bar3]}^{(S)} = -\frac{5}{6}\,h_{4,[66]}^{(S)}- \frac{4}{3}\,h_{5,[66]}^{(S)}\,,
\nonumber\\
&&  g_{0,[66]}^{(S)} = g_{1,[\bar 36]}^{(S)}= g_{D,[\bar 36]}^{(S)}= 0\,,\;\;\;\;\, \qquad\quad
g_{1,[66]}^{(S)} = -h_{4,[66]}^{(S)} -\frac{2}{3}\,h_{5,[66]}^{(S)}\,, \qquad \qquad 
g_{D,[66]}^{(S)}= - h_{2,[66]}^{(S)}\,,
\nonumber\\
&& h_{0,[66]}^{(S)} = 
h_{1,[66]}^{(S)} = h_{3,[66]}^{(S)} = 0\,,
\nonumber\\ \nonumber\\ 
&& g_{0,[\bar 3\bar 3]}^{(V)} = -\frac{11}{3}\,h_{1,[66]}^{(V)}+ \frac{4}{3}\,h_{2,[66]}^{(V)}\,, \qquad \qquad \,
g_{1,[\bar 3\bar 3]}^{(V)}=0  \,,\qquad \qquad
g_{D,[\bar 3\bar 3]}^{(V)} = 6\,h_{1,[66]}^{(V)}- 3\,h_{2,[66]}^{(V)}\,,
\nonumber\\
&& g_{1,[\bar36]}^{(V)} = g_{D,[\bar36]}^{(V)} =g_{0,[66]}^{(V)}  = h_{0,[66]}^{(V)} = 0\,,\qquad\qquad
g_{1,[66]}^{(V)} = -\,h_{1,[66]}^{(V)}\,,\qquad\qquad
 g_{D,[66]}^{(V)} =- h_{2,[66]}^{(V)}\,,
\nonumber\\ \nonumber\\ 
&& f_{1,[\bar36]}^{(A)} = \sqrt{3}\,g_{1,[\bar36]}^{(T)} \,,\qquad \qquad \;\;
f_{F,[\bar36]}^{(A)} = -2\,h_{F,[66]}^{(T)}\,,\qquad \qquad f_{D,[\bar 36]}^{(A)} = f_{0,[66]}^{(A)} = f_{D,[66]}^{(A)} = 0\,,  
\nonumber\\
&& f_{1,[66]}^{(A)} = \frac{1}{\sqrt{3}}\,h_{5,[66]}^{(S)}\,,\qquad\qquad \,
f_{F,[66]}^{(A)} = - \frac{2}{\sqrt{3}}\,h_{F,[66]}^{(T)}\,,
\nonumber\\
&& g_{F,[\bar 3 \bar 3]}^{(T)} = 0\,, \qquad \qquad \qquad \quad \;\;\, 
g_{F,[\bar 3 6]}^{(T)} = \frac{1}{\sqrt{3}}\,h_{F,[66]}^{(T)}\,,
\qquad \qquad 
 g_{F,[66]}^{(T)} = -\frac{2}{3}\,h_{F,[66]}^{(T)}\,.
\label{large-Nc sumrules}
\end{eqnarray}
It is interesting to compare our large-$N_c$ sum rules (\ref{large-Nc sumrules}) with
the sum rules implied by the heavy-quark spin symmetry in (\ref{HQS-baryon sumrules}).
The heavy-quark spin operator $J_Q^i$, which would violate the spin-symmetry
explicitly, contributes at subleading orders in the operator expansion of (\ref{def-AA-Nc}) only.
Therefore one may expect that the  spin-symmetry sum rules in (\ref{HQS-baryon sumrules}) do not provide
additional constraints on the parameters. However, this is not the case. The combination of (\ref{large-Nc sumrules})
with the spin-symmetry sum rules (\ref{HQS-baryon sumrules}) does lead to one extra relation, that is implied by
\begin{eqnarray}
g_6 + g_7 = 0 \qquad \leftrightarrow \qquad f^{(A)}_{1,[66]} = 0 \,.
\end{eqnarray}
This does not contradict the systematics of the large-$N_c$ operator expansion. It merely shows that
the coefficient $g_6+ g_7$ is  
suppressed in the heavy-quark mass expansion.
Though the operator analysis is not predicting such a feature, it can not exclude it.

\newpage

\section{Summary}

We constructed the chiral $SU(3)$ Lagrangian with charmed baryons with spin $J^P=1/2^+$ and $J^P=3/2^+$ at subleading orders
as it is required for a chiral extrapolation of the charmed baryon masses from lattice QCD simulations. All
counter terms that are relevant at next-to-next-to-next-to-leading order (N$^3$LO) for such a chiral extrapolation
were identified.

\begin{itemize}

\item[$\bullet$]

At N$^2$LO we find 16 low-energy parameters. There are 3 mass parameters for the
anti-triplet and the two sextet baryons, 6 parameters describing the meson-baryon vertices and 7 symmetry breaking
parameters. The heavy-quark spin symmetry predicts four sum rules for the meson-baryon vertices and degenerate masses
for the two baryon sextet fields. Here a large-$N_c$ operator analysis at NLO suggests the relevance of one further
spin-symmetry breaking parameter.

\item[$\bullet $]
At N$^3$LO there are additional 17 chiral symmetry breaking parameters
and 24 symmetry preserving parameters. For the leading symmetry conserving two-body counter terms
involving two baryon fields and two Goldstone boson fields we find 36 terms. While the heavy-quark spin symmetry leads to
$36-16=20$ sum rules, an expansion in $1/N_c$ at next-to-leading order (NLO) generates $36-7 = 29$ parameter relations.
A combined expansion leaves 6 unknown parameters only, a parameter reduction by about a factor of 10.
For the symmetry breaking counter terms we find 17 terms, for which there are $17-9 = 8$ sum rules from the heavy-quark
spin symmetry and $17-5 = 12 $ sum rules from a $1/N_c$ expansion at NLO.  Here a combined expansion does not further reduce
the number of parameters.
\end{itemize}

At present such sum rules can not be confronted directly with empirical information. They are useful constraints in
establishing a systematic coupled-channel effective field theory for the Goldstone-boson charmed-baryon scattering
beyond the threshold region. In the near future a significant increase of the lattice data base on charmed baryon
masses is expected. They will allow a significant test of our results.

\vskip0.3cm
{\bfseries{Acknowledgments}}
\vskip0.3cm
Daris Samart was supported by Thailand reseach fund TRF-RMUTI under contract No. TRG5680079.

\newpage
\section{Appendix}

We consider matrix elements of the symmetric product of two one-body operators ${\mathcal O}$ in the
charmed baryon ground state at $N_c =3$. The generic notation
\begin{eqnarray}
&&\langle\, {\mathcal O} \,\rangle^{\pm \pm}_{\, \bar SS} \equiv \;
\roundbra{\,mn_\pm,\, \bar S, \,\bar\chi \,} \,{\mathcal O}\,\roundket{\,kl_{\pm},\,S,\,\chi\,}\,,
\nonumber\\
&& \langle\, {\mathcal O} \,\rangle^{\pm \mp}_{\,\bar SS} \equiv \;
\roundbra{\,mn_\pm,\, \bar S, \,\bar\chi \,} \,{\mathcal O}\,\roundket{\,kl_{\mp},\,S,\,\chi\,}\,,
\end{eqnarray}
will be applied. The results are expressed in terms of the flavour structures $\Lambda_{(kl)_+}^{(a),\,(rs)_\pm}$
and $\Lambda_{(kl)_-}^{(a),\,(rs)_\pm}$ and the spin structures $\sigma_i $ and $S_i$ introduced
in (\ref{def:spin-transition-matrices}). The following identities turn useful
\begin{eqnarray}
&& \delta_{(ij)_\pm}^{(mn)_\pm}
=  \frac 12\,\Big( \delta_{mi}\,\delta_{nj} \pm \delta_{ni}\,\delta_{mj} \,\Big)\,,
\nonumber\\
&& \Lambda_{(kl)_+}^{(a),\,(rs)_\pm}\,\delta_{(rs)_\pm}^{(mn)_\pm}
= \Lambda_{(kl)_+}^{(a),\,(mn)_\pm}\,, \qquad
\Lambda_{(kl)_-}^{(a),\,(rs)_\pm}\,\delta_{(rs)_\pm}^{(mn)_\pm}
= \Lambda_{(kl)_-}^{(a),\,(mn)_\pm}\,,
\nonumber\\ \nonumber\\
&&\Lambda_{(rs)_-}^{a\,,\,(mn)_+}\,\Lambda_{(kl)_+}^{b\,,\,(rs)_-} + \Lambda_{(rs)_-}^{b\,,\,(mn)_+}\,\Lambda_{(kl)_+}^{a\,,\,(rs)_-}
= -\Lambda_{(rs)_+}^{a\,,\,(mn)_+}\,\Lambda_{(kl)_+}^{b\,,\,(rs)_+}
- \Lambda_{(rs)_+}^{b\,,\,(mn)_+}\,\Lambda_{(kl)_+}^{a\,,\,(rs)_+}
\nonumber\\
&& \qquad \qquad \qquad +\, \frac 43\,\delta^{ab}\,\delta_{(kl)_+}^{(mn)_+} + 2\,d^{abc}\,\Lambda_{(kl)_+}^{c\,,\,(mn)_+} \,,
\nonumber\\
&& \Lambda_{(rs)_+}^{(a),\,(mn)_-}\,\Lambda_{(kl)_-}^{(b),\,(rs)_+} + \Lambda_{(rs)_+}^{(b),\,(mn)_-}\,\Lambda_{(kl)_-}^{(a),\,(rs)_+}
= \delta_{ab}\,\delta^{(mn)_-}_{(kl)_-}
 +3\, d_{abc}\,\Lambda^{c,(mn)_-}_{(kl)_-} \,,
\nonumber\\
&& \Lambda_{(rs)_-}^{(a),\,(mn)_-}\,\Lambda_{(kl)_-}^{(b),\,(rs)_-} + \Lambda_{(rs)_-}^{(b),\,(mn)_-}\,\Lambda_{(kl)_-}^{(a),\,(rs)_-}
= \frac{1}{3}\,\delta_{ab}\,\delta^{(mn)_-}_{(kl)_-}
 -d_{abc}\,\Lambda^{c,(mn)_-}_{(kl)_-} \,,
 \nonumber\\
&& \Lambda_{(rs)_\pm}^{(a),\,(mn)_-}\,\Lambda_{(kl)_+}^{(b),\,(rs)_\pm} + \Lambda_{(rs)_\pm}^{(b),\,(mn)_-}\,\Lambda_{(kl)_+}^{(a),\,(rs)_\pm}
= d_{abc}\,\Lambda^{c,(mn)_-}_{(kl)_+} \,,
 \nonumber\\
&& \Lambda_{(rs)_\pm}^{(a),\,(mn)_+}\,\Lambda_{(kl)_-}^{(b),\,(rs)_\pm} + \Lambda_{(rs)_\pm}^{(b),\,(mn)_+}\,\Lambda_{(kl)_-}^{(a),\,(rs)_\pm}
= d_{abc}\,\Lambda^{c,(mn)_+}_{(kl)_-} \,,
\nonumber\\ \nonumber\\
&& \Lambda_{(rs)_\pm}^{a\,,\,(mn)_+}\,\Lambda_{(kl)_+}^{b\,,\,(rs)_\pm} - \Lambda_{(rs)_\pm}^{b\,,\,(mn)_+}\,\Lambda_{(kl)_+}^{a\,,\,(rs)_\pm} = i\,f_{abc}\,\Lambda_{(kl)_+}^{a\,,\,(mn)_+}\,,
\nonumber\\
&& \Lambda_{(rs)_\pm}^{a\,,\,(mn)_-}\,\Lambda_{(kl)_-}^{b\,,\,(rs)_\pm} - \Lambda_{(rs)_\pm}^{b\,,\,(mn)_-}\,\Lambda_{(kl)_-}^{a\,,\,(rs)_\pm} = i\,f_{abc}\,\Lambda_{(kl)_-}^{a\,,\,(mn)_-}
\,,
\nonumber\\
&& \Lambda_{(rs)_+}^{a\,,\,(mn)_+}\,\Lambda_{(kl)_-}^{b\,,\,(rs)_+}
+ \Lambda_{(rs)_-}^{a\,,\,(mn)_+}\,\Lambda_{(kl)_-}^{b\,,\,(rs)_-} - (a \leftrightarrow b)
= 2\,i\,f_{abc}\,\Lambda_{(kl)_-}^{a\,,\,(mn)_+}\,,
\nonumber\\
&& \Lambda_{(rs)_+}^{a\,,\,(mn)_-}\,\Lambda_{(kl)_+}^{b\,,\,(rs)_+}
+ \Lambda_{(rs)_-}^{a\,,\,(mn)_-}\,\Lambda_{(kl)_+}^{b\,,\,(rs)_-} - (a \leftrightarrow b)
= 2\,i\,f_{abc}\,\Lambda_{(kl)_+}^{a\,,\,(mn)_ -}\,.
\end{eqnarray}
We find
\allowdisplaybreaks
\begin{eqnarray}
&& \langle\,\big\{\,J_i,\,J_j\,\big\}\,\rangle^{++}_{\frac{1}{2}\frac{1}{2}}
= \frac{4}{3}\,\delta_{ij}\,\delta_{\bar\chi\,\chi}\,\delta_{(kl)_+}^{(mn)_+}\,, \qquad
 \langle\,\big\{\,J_i,\,T^a\,\big\}\,\rangle^{++}_{\frac{1}{2}\frac{1}{2}}
= \frac 43\,\sigma_{\bar\chi\chi}^{(i)}\,\Lambda_{(kl)_+}^{(a),\,(mn)_+}\,,
\nonumber\\
&& \langle\,\big\{\,J_i,\,G^{a}_j\,\big\}\,\rangle^{++}_{\frac{1}{2}\frac{1}{2}}
= \frac{2}{3}\,\delta_{ij}\,\delta_{\bar\chi\,\chi}\,\Lambda_{(kl)_+}^{(a),\,(mn)_+}\,,
\nonumber\\
&& \langle\,\big\{\,T^a,\,T^b\,\big\}\,\rangle^{++}_{\frac{1}{2}\frac{1}{2}}
= \textcolor{black}{\Big(\Lambda_{(rs)_+}^{(a),\,(mn)_+}\,\Lambda_{(kl)_+}^{(b),\,(rs)_+}
+ \Lambda_{(rs)_+}^{(b),\,(mn)_+}\,\Lambda_{(kl)_+}^{(a),\,(rs)_+}
\,\Big)\,\delta_{\bar\chi\chi} }\,,
\nonumber\\
&& \langle\,\big\{\,T^a,\,G^{b}_i\,\big\}\,\rangle^{++}_{\frac{1}{2}\frac{1}{2}}
= \frac 13\,\sigma_{\bar\chi\chi}^{(i)}\,\Big(\Lambda_{(rs)_+}^{(a),\,(mn)_+}\,\Lambda_{(kl)_+}^{(b),\,(rs)_+}
+ \Lambda_{(rs)_+}^{(b),\,(mn)_+}\,\Lambda_{(kl)_+}^{(a),\,(rs)_+}\,\Big)\,,
\nonumber\\
&& \langle\,\big\{\,G^{a}_i,\,G^{b}_j\,\big\}\,\rangle^{++}_{\frac{1}{2}\frac{1}{2}}
= \frac 16\,\delta_{ij}\,\delta_{\bar\chi\chi}
\,\Big(\Lambda_{(rs)_+}^{(a),\,(mn)_+}\,\Lambda_{(kl)_+}^{(b),\,(rs)_+}\,
+ (a \leftrightarrow b) \,\Big)
\nonumber\\
&& \qquad \quad +\, \frac{1}{12}\,i\,\epsilon_{ijk}\,\sigma_{\bar\chi\chi}^{(k)}
\,\Big(\Lambda_{(rs)_+}^{(a),\,(mn)_+}\,\Lambda_{(kl)_+}^{(b),\,(rs)_+}\,
- (a \leftrightarrow b) \,\Big)
\nonumber\\
&& \qquad \quad +\, \frac{1}{12}\,\delta_{ij}\,\delta_{\bar\chi\chi}
\,\Big(\Lambda_{(rs)_-}^{(a),\,(mn)_+}\,\Lambda_{(kl)_+}^{(b),\,(rs)_-}
+ (a \leftrightarrow b)\,\Big)
\nonumber\\
&& \qquad \quad +\,\frac{1}{12}\,i\,\epsilon_{ijk}\,\sigma_{\bar\chi\chi}^{(k)}
\,\Big(\Lambda_{(rs)_-}^{(a),\,(mn)_+}\,\Lambda_{(kl)_+}^{(b),\,(rs)_-}
- (a \leftrightarrow b)\,\Big)\,,
\nonumber\\
&& \langle\,\big\{\,J_Q^i,\,J_j\,\big\}\,\rangle^{++}_{\frac{1}{2}\frac{1}{2}}
= \frac{2}{3}\,\delta_{ij}\,\delta_{\bar\chi\,\chi}\,\delta_{(kl)_+}^{(mn)_+}\,, \qquad
 \langle\,\big\{\,J_Q^i,\,J_Q^j\,\big\}\,\rangle^{++}_{\frac{1}{2}\frac{1}{2}}
= \frac 12\,\delta_{ij}\,\delta_{\bar\chi\,\chi}\,\delta_{(kl)_+}^{(mn)_+}\,,
\nonumber\\
&& \langle\,\big\{\,J_Q^i,\,T^a\,\big\}\,\rangle^{++}_{\frac{1}{2}\frac{1}{2}}
= \sigma_{\bar\chi\chi}^{(i)}\,\Lambda_{(kl)_+}^{(a),\,(mn)_+}\,, \qquad \;\,
 \langle\,\big\{\,J_Q^i,\,G^{a}_j\,\big\}\,\rangle^{++}_{\frac{1}{2}\frac{1}{2}}
= \frac{1}{3}\,\delta_{ij}\,\delta_{\bar\chi\,\chi}\,\Lambda_{(kl)_+}^{(a),\,(mn)_+}\,,
\nonumber\\ \nonumber\\
&& \langle\,\big\{\,J_i,\,J_j\,\big\}\,\rangle^{++}_{\frac{3}{2} \frac{3}{2}}
=\Big( \textcolor{black}{ 2\,\delta_{ij}
- \,S_i\,S_j^{\dagger} - S_j\,S_i^{\dagger} } \Big)_{\bar\chi\chi}\,\delta_{(kl)_+}^{(mn)_+}\,,
\nonumber\\
&& \langle\,\big\{\,J_i,\,T^a\,\big\}\,\rangle^{++}_{\frac{3}{2} \frac{3}{2}}
= 2\,(\vec S\,\sigma_i\,\vec S^\dagger)_{\bar\chi\chi}\,\Lambda_{(kl)_+}^{(a),\,(mn)_+}\,,
\nonumber\\
&& \langle\,\big\{\,J_i,\,G^{a}_j\,\big\}\,\rangle^{++}_{\frac{3}{2} \frac{3}{2}}
= \frac 12\,\Big( \textcolor{black}{2\,\delta_{ij}\
 -S_i\,S_j^{\dagger} - S_j\,S_i^{\dagger} }\Big)_{\bar\chi\chi}
\,\Lambda_{(kl)_+}^{(a),\,(mn)_+}\,,
\nonumber\\
&& \langle\,\big\{\,T^a,\,T^b\,\big\}\,\rangle^{++}_{\frac{3}{2} \frac{3}{2}}
= \textcolor{black}{ \Big(\Lambda_{(rs)_+}^{(a),\,(mn)_+}\,\Lambda_{(kl)_+}^{(b),\,(rs)_+}
+ \Lambda_{(rs)_+}^{(b),\,(mn)_+}\,\Lambda_{(kl)_+}^{(a),\,(rs)_+}
\,\Big)\,}\delta_{\bar\chi\chi}\,,
\nonumber\\
&& \langle\,\big\{\,T^a,\,G^{b}_i\,\big\}\,\rangle^{++}_{\frac{3}{2} \frac{3}{2}}
= \frac 12\,(\vec S\,\sigma_i\,\vec S^\dagger)_{\bar\chi\chi}\,\Big(\Lambda_{(rs)_+}^{(a),\,(mn)_+}\,\Lambda_{(kl)_+}^{(b),\,(rs)_+}+ \Lambda_{(rs)_+}^{(b),\,(mn)_+}\,\Lambda_{(kl)_+}^{(a),\,(rs)_+}\,\Big)\,,
\nonumber\\
&& \langle\,\big\{\,G^{a}_i,\,G^{b}_j\,\big\}\,\rangle^{++}_{\frac{3}{2} \frac{3}{2}}
=\textcolor{black}{ \frac{1}{4}}\,\delta_{ij}\,\delta_{\bar\chi\chi}\,\Big(\Lambda_{(rs)_+}^{(a),\,(mn)_+}\,\Lambda_{(kl)_+}^{(b),\,(rs)_+}
+ (a \leftrightarrow b)\,\Big)
\nonumber\\
&& \qquad \quad +\, \textcolor{black}{ \frac{1}{8}}\,i\,\epsilon_{ijk}\,(\vec S\,\sigma^{(k)}\,\vec S^\dagger)_{\bar\chi\chi}
\,\Big(\Lambda_{(rs)_+}^{(a),\,(mn)_+}\,\Lambda_{(kl)_+}^{(b),\,(rs)_+}
- (a \leftrightarrow b)\,\Big)
\nonumber\\
&& \qquad \quad -\,\textcolor{black}{ \frac{1}{8}}\,\big( S_i\,S_j^{\dagger} + S_j\,S_i^{\dagger} \big)_{\bar\chi\chi}
\,\Big(\Lambda_{(rs)_+}^{(a),\,(mn)_+}\,\Lambda_{(kl)_+}^{(b),\,(rs)_+}
+ (a \leftrightarrow b)\,\Big)
\nonumber\\
&& \qquad \quad +\, \textcolor{black}{ \frac{1}{8}}\,i\,\epsilon_{ijk}\,(\vec S\,\sigma^{(k)}\,\vec S^\dagger)_{\bar\chi\chi}
\,\Big(\Lambda_{(rs)_-}^{(a),\,(mn)_+}\,\Lambda_{(kl)_+}^{(b),\,(rs)_-}
- (a \leftrightarrow b)\,\Big)
\nonumber\\
&& \qquad \quad +\, \textcolor{black}{ \frac{1}{8}}\,\big( S_i\,S_j^{\dagger} + S_j\,S_i^{\dagger} \big)_{\bar\chi\chi}
\,\Big(\Lambda_{(rs)_-}^{(a),\,(mn)_+}\,\Lambda_{(kl)_+}^{(b),\,(rs)_-}
+ (a \leftrightarrow b)\,\Big) \,,
\nonumber\\
&& \langle\,\big\{\,J_Q^i,\,J_j\,\big\}\,\rangle^{++}_{\frac{3}{2} \frac{3}{2}}
= \Big( \textcolor{black}{\delta_{ij}- S_i\,S_j^{\dagger} - S_j\,S_i^{\dagger} }
\Big)_{\bar\chi\chi}\,\delta_{(kl)_+}^{(mn)_+}\,, \qquad
 \langle\,\big\{\,J_Q^i,\,J_Q^j\,\big\}\,\rangle^{++}_{\frac{3}{2} \frac{3}{2}}
= \textcolor{black}{\frac{1}{2}\,\delta_{ij}\,\delta_{\bar\chi\,\chi}\,\delta_{(kl)_+}^{(mn)_+}}\,,
\nonumber\\
&& \langle\,\big\{\,J_Q^i,\,T^a\,\big\}\,\rangle^{++}_{\frac{3}{2} \frac{3}{2}}
= (\vec S\,\sigma^{(i)}\,\vec S^\dagger)_{\bar\chi\chi}\,\Lambda_{(kl)_+}^{(a),\,(mn)_+}\,,
\nonumber\\
&& \langle\,\big\{\,J_Q^i,\,G^{a}_j\,\big\}\,\rangle^{++}_{\frac{3}{2} \frac{3}{2}}
= \frac{1}{2}\, \Big( \textcolor{black}{\delta_{ij}
 -\,S_i\,S_j^{\dagger} - S_j\,S_i^{\dagger} }
\Big)_{\bar\chi\chi}\,\Lambda_{(kl)_+}^{(a),\,(mn)_+}\,,
\nonumber\\ \nonumber\\
&& \langle\,\big\{\,J_i,\,J_j\,\big\}\,\rangle^{--}_{\frac{1}{2} \frac{1}{2}}
= 0\,, \qquad \;\;
 \langle\,\big\{\,J_i,\,T^a\,\big\}\,\rangle^{--}_{\frac{1}{2} \frac{1}{2}}
= 0\,, \qquad
 \langle\,\big\{\,J_i,\,G^{b}_j\,\big\}\,\rangle^{--}_{\frac{1}{2} \frac{1}{2}}
= 0\,,
\nonumber\\
&& \langle\,\big\{\,T^a,\,T^b\,\big\}\,\rangle^{--}_{\frac{1}{2} \frac{1}{2}}
= \Big(\textcolor{black}{\Lambda_{(rs)_-}^{(a),\,(mn)_-}\,\Lambda_{(kl)_-}^{(b),\,(rs)_-}
+ \Lambda_{(rs)_-}^{(b),\,(mn)_-}\,\Lambda_{(kl)_-}^{(a),\,(rs)_-} }
\,\Big)\,\delta_{\bar\chi\chi}\,,
\nonumber\\
&& \langle\,\big\{\,T^a,\,G^{b}_i\,\big\}\,\rangle^{--}_{\frac{1}{2} \frac{1}{2}}
= 0\,, \qquad  \langle\,\big\{\,J_Q^i,\,J_j\,\big\}\,\rangle^{--}_{\frac{1}{2} \frac{1}{2}}
= 0\,, \qquad \langle\,\big\{\,J_Q^i,\,G^{a}_j\,\big\}\,\rangle^{--}_{\frac{1}{2} \frac{1}{2}}
= 0\,,
\nonumber\\
&& \langle\,\big\{\,G^{a}_i,\,G^{b}_j\,\big\}\,\rangle^{--}_{\frac{1}{2} \frac{1}{2}}
= \frac{1}{4}\,\delta_{ij}\,\delta_{\bar\chi\,\chi}\,\Big( \,\Lambda_{(rs)_+}^{(a),\,(mn)_-}
\,\Lambda_{(kl)_-}^{(b),\,(rs)_+} + \Lambda_{(rs)_+}^{(b),\,(mn)_-}
\,\Lambda_{(kl)_-}^{(a),\,(rs)_+}\,\Big)\,,
\nonumber\\
&& \langle\,\big\{\,J_Q^i,\,J_Q^j\,\big\}\,\rangle^{--}_{\frac{1}{2} \frac{1}{2}}
= \frac 12\,\delta_{ij}\,\delta_{\bar\chi\,\chi}\,\delta_{(kl)_-}^{(mn)_-}\,, \qquad
 \langle\,\big\{\,J_Q^i,\,T^a\,\big\}\,\rangle^{--}_{\frac{1}{2} \frac{1}{2}}
= \sigma_{\bar\chi\chi}^{(i)}\,\Lambda_{(kl)_-}^{(a),\,(mn)_-}\,,
\nonumber\\ \nonumber\\
&& \langle\,\big\{\,J_i,\,J_j\,\big\}\,\rangle^{++}_{\frac{3}{2} \frac{1}{2}}
= -\frac{1}{\sqrt{3}}\,\Big(\,S_i\,\sigma_j + S_j\,\sigma_i\Big)_{\bar\chi\chi}\,\delta_{(kl)_+}^{(mn)_+}\,,
\nonumber\\
&& \langle\,\big\{\,J_i,\,T^a\,\big\}\,\rangle^{++}_{\frac{3}{2} \frac{1}{2}}
= -\frac{2}{\sqrt 3}\,S_{\bar\chi\chi}^{(i)}\,\,\Lambda_{(kl)_+}^{(a),\,(mn)_+}\,,\qquad
\textcolor{black}{\langle\,\big\{\,J^{(i)}_Q,\,T^a\,\big\}\,\rangle^{++}_{\frac{3}{2} \frac{1}{2}}
= \frac{1}{\sqrt 3}\,S_{\bar\chi\chi}^{(i)}\,\,\Lambda_{(kl)_+}^{(a),\,(mn)_+}}\,,
\nonumber\\
&& \langle\,\big\{\,J_i,\,G^{a}_j\,\big\}\,\rangle^{++}_{\frac{3}{2} \frac{1}{2}}
= -\,\frac{1}{2\sqrt{3}}\,\Big(\,S_i\,\sigma_j + S_j\,\sigma_i \Big)_{\bar\chi\chi}
\,\Lambda_{(kl)_+}^{(a),\,(mn)_+}\,,
\nonumber\\
&& \langle\,\big\{\,T^a,\,G^{b}_i\,\big\}\,\rangle^{++}_{\frac{3}{2} \frac{1}{2}}
= -\frac{1}{2\sqrt 3}\,S_{\bar\chi\chi}^{(i)}\,\Big(\Lambda_{(rs)_+}^{(a),\,(mn)_+}\,\Lambda_{(kl)_+}^{(b),\,(rs)_+}
+\Lambda_{\,(rs)_+}^{(b),\,(mn)_+}\,\Lambda_{(kl)_+}^{(a),(rs)_+}
\,\Big)\,,
\nonumber\\
&& \langle\,\big\{\,G^{a}_i,\,G^{b}_j\,\big\}\,\rangle^{++}_{\frac{3}{2} \frac{1}{2}}
= -\frac{1}{8\sqrt{3}}\,i\,\epsilon_{ijk}\,S_{\bar\chi\chi}^{(k)}\,\Big(
\Lambda_{(rs)_+}^{(a),\,(mn)_+}\,\Lambda_{(kl)_+}^{(b),\,(rs)_+}
- (a \leftrightarrow b)\,\Big)
\nonumber\\
&& \qquad \quad  -\frac{1}{8\sqrt{3}}\,\Big(\,S_i\,\sigma_j + S_j\,\sigma_i\Big)_{\bar\chi\chi}\,
\Big(\Lambda_{(rs)_+}^{(a),\,(mn)_+}\,\Lambda_{(kl)_+}^{(b),\,(rs)_+}
+ (a \leftrightarrow b)\,\Big)
\nonumber\\
&& \qquad \quad -\frac{1}{8\sqrt{3}}\,i\,\epsilon_{ijk}\,S_{\bar\chi\chi}^{(k)}\,\Big(
\Lambda_{(rs)_-}^{(a),\,(mn)_+}\,\Lambda_{(kl)_+}^{(b),\,(rs)_-}
- (a \leftrightarrow b)\,\Big)
\nonumber\\
&& \qquad \quad +\,\frac{1}{8\sqrt{3}}\,\big(S_i\,\sigma_j + S_j\,\sigma_i\big)_{\bar\chi\chi}
\,\Big(
\Lambda_{(rs)_-}^{(a),\,(mn)_+}\,\Lambda_{(kl)_+}^{(b),\,(rs)_-}
+ (a \leftrightarrow b)\,\Big)\,,
\nonumber\\
&& \langle\,\big\{\,J_Q^i,\,J_j\,\big\}\,\rangle^{++}_{\frac{3}{2} \frac{1}{2}}
= -\frac{1}{\sqrt{3}}\,i\,\epsilon_{ijk}\,S_{\bar\chi\chi}^{(k)}
\,\delta_{(kl)_+}^{(mn)_+}\,, \qquad
 \langle\,\big\{\,J_Q^i,\,G^{a}_j\,\big\}\,\rangle^{++}_{\frac{3}{2} \frac{1}{2}}
= -\frac{1}{2\,\sqrt{3}}\,i\,\epsilon_{ijk}\,S_{\bar\chi\chi}^{(k)}\,\Lambda_{(kl)_+}^{(a),\,(mn)_+}\,,
\nonumber\\ \nonumber\\
&& \langle\,\big\{\,J_i,\,J_j\,\big\}\,\rangle^{+-}_{\frac{3}{2} \frac{1}{2}}
= 0\,, \qquad
 \langle\,\big\{\,J_i,\,T^a\,\big\}\,\rangle^{+-}_{\frac{3}{2} \frac{1}{2}}
= 0\,,\qquad \langle\,\big\{\,J_Q^i,\,J_j\,\big\}\,\rangle^{+-}_{\frac{3}{2} \frac{1}{2}}
= 0\,,
\nonumber\\
&& \langle\,\big\{\,J_i,\,G^{a}_j\,\big\}\,\rangle^{+-}_{\frac{3}{2} \frac{1}{2}}
= -\frac{1}{2}\,i\,\epsilon_{ijk}\,S_{\bar\chi\chi}^{(k)}
\,\Lambda_{(kl)_-}^{(a),\,(mn)_+}\,, \qquad \quad \,
\textcolor{black}{\langle\,\big\{\,J_Q^i,\,T^a\,\big\}\,\rangle^{+-}_{\frac{3}{2} \frac{1}{2}}
= 0\,,}
\nonumber\\
&& \langle\,\big\{\,T^a,\,G^{b}_i\,\big\}\,\rangle^{+-}_{\frac{3}{2} \frac{1}{2}}
= -\,\frac 12\,S_{\bar\chi\chi}^{(i)}\,\Big(\Lambda_{(rs)_+}^{(a),\,(mn)_+}\,\Lambda_{(kl)_-}^{(b),\,(rs)_+}
+ \Lambda_{(rs)_-}^{(b),\,(mn)_+}\,\Lambda_{(kl)_-}^{(a),\,(rs)_-}
\,\Big)\,,
\nonumber\\
&& \langle\,\big\{\,G^{a}_i,\,G^{b}_j\,\big\}\,\rangle^{+-}_{\frac{3}{2} \frac{1}{2}}
= - \,\frac{1}{4}\,i\,\epsilon_{ijk}\,S_{\bar\chi\chi}^{(k)}\,\Big(
\Lambda_{(rs)_+}^{(a),\,(mn)_+}\,\Lambda_{(kl)_-}^{(b),\,(rs)_+}
- \Lambda_{(rs)_+}^{(b),\,(mn)_+}\,\Lambda_{(kl)_-}^{(a),\,(rs)_+}\,\Big)\,,
\nonumber\\
&& \langle\,\big\{\,J_Q^i,\,G^{a}_j\,\big\}\,\rangle^{+-}_{\frac{3}{2} \frac{1}{2}}
= -\, \frac{1}{4}\,\Big( i\,\epsilon_{ijk}\,S_{\bar\chi\chi}^{(k)}
+ \big(S_i\,\sigma_j+ S_j\,\sigma_i\big)_{\bar\chi\chi}\,\Big)\,\Lambda_{(kl)_-}^{(a),\,(mn)_+} \,,
\nonumber\\ \nonumber\\
&& \langle\,\big\{\,J_i,\,J_j\,\big\}\,\rangle^{-+}_{\frac{1}{2} \frac{1}{2}}
= 0\,, \qquad
 \langle\,\big\{\,J_i,\,T^a\,\big\}\,\rangle^{-+}_{\frac{1}{2} \frac{1}{2}}
= 0\,, \qquad  \langle\,\big\{\,J_Q^i,\,J_j\,\big\}\,\rangle^{-+}_{\frac{1}{2} \frac{1}{2}}
= 0\,,
\nonumber\\
&& \langle\,\big\{\,J_i,\,G^{a}_j\,\big\}\,\rangle^{-+}_{\frac{1}{2} \frac{1}{2}}
= \frac{1}{2\sqrt{3}}\,i\,\epsilon_{ijk}\,\sigma_{\bar\chi\chi}^{(k)}
\,\Lambda_{(kl)_+}^{(a),\,(mn)_-}\,,
\nonumber\\
&& \langle\,\big\{\,T^a,\,G^{b}_i\,\big\}\,\rangle^{-+}_{\frac{1}{2} \frac{1}{2}}
= -\frac{1}{2\sqrt 3}\,\sigma_{\bar\chi\chi}^{(i)}\,\Big(\Lambda_{(rs)_-}^{(a),\,(mn)_-}\,\Lambda_{(kl)_+}^{(b),\,(rs)_-} + \Lambda_{(rs)_+}^{(b),\,(mn)_-}\,\Lambda_{(kl)_+}^{(a),\,(rs)_+}
\,\Big)\,,
\nonumber\\
&& \langle\,\big\{\,G^{a}_i,\,G^{b}_j\,\big\}\,\rangle^{-+}_{\frac{1}{2} \frac{1}{2}}
= -\frac{1}{4\sqrt{3}}\,i\,\epsilon_{ijk}\,\sigma_{\bar\chi\chi}^{(k)}\,\Big(
\Lambda_{(rs)_+}^{(a),\,(mn)_-}\,\Lambda_{(kl)_+}^{(b),\,(rs)_+}
- \Lambda_{(rs)_+}^{(b),\,(mn)_-}\,\Lambda_{(kl)_+}^{(a),\,(rs)_+}\,\Big)\,,
\nonumber\\
&& \langle\,\big\{\,J_Q^i,\,G^{a}_j\,\big\}\,\rangle^{-+}_{\frac{1}{2} \frac{1}{2}}
= -\,\frac{1}{2\sqrt{3}}\,\delta_{ij}\,\delta_{\bar\chi\chi}
\,\Lambda_{(kl)_+}^{(a),\,(mn)_-}\,.
\end{eqnarray}
Note that all results of this Appendix are
valid also for the singlet components with $a=0$ or $b=0$ if the convention
(\ref{def-redundant-operator}) is used together with $\lambda_0 = \sqrt{2/3} \,\one$  in (\ref{def:spin-transition-matrices}).

\newpage

\bibliography{1}

\end{document}